\documentclass[twocolumn,showpacs,preprintnumbers,amsmath,amssymb,prd]{revtex4}

\usepackage{graphicx}
\usepackage{rotating}
\usepackage{color}

\newcommand{\E}[1]{ \cdot 10^{#1}}
\newcommand{\RO}{{\bf{\em R}}}

\begin{document}
\title{ Analysis of 3 years of data from the gravitational wave detectors \\
EXPLORER and NAUTILUS}
\author{
%Lista degli autori
P. Astone$^1$, M. Bassan$^2$
%\footnote{*}{Corresponding author: massimo.bassan@roma2.infn.it}
, E. Coccia$^2$, S. D'Antonio$^2$,\\
V. Fafone$^2$, 
G. Giordano$^3$, A. Marini$^3$, Y. Minenkov$^2$,\\
I. Modena$^2$, A. Moleti$^2$, G. V. Pallottino$^1$, G. Pizzella$^3$,\\
A. Rocchi$^2$, F. Ronga$^3$, R. Terenzi$^4$, M. Visco$^4$.\\
$~$\\
{\it ${}^{1)}$ University of Rome "La Sapienza" and INFN, Rome1}\\
{\it ${}^{2)}$ University of Rome "Tor Vergata" and INFN, Rome2}\\
{\it ${}^{3)}$ Istituto Nazionale di Fisica Nucleare INFN, LNF}\\
{\it ${}^{4)}$ IAPS-INAF and INFN, Rome2}\
}

\begin{abstract}

We performed a search for short gravitational wave bursts using
about 3 years of data of the resonant bar detectors Nautilus and
Explorer. Two types of analysis were performed:
a search for coincidences with a low background
of accidentals (0.1 over the entire period), and the
calculation of upper limits on the rate of gravitational wave bursts.
Here we give a detailed account of the
methodology and we report the results:  a null search for 
coincident events and an upper limit that improves over all 
previous limits from resonant antennas, and  is competitive, 
in the range $h_{rss}\sim 10^{-19}$, with limits from 
interferometric detectors.
Some new methodological features are introduced that have proven
successful in the upper limits evaluation.
\end{abstract}
\maketitle

\section{Introduction}

In the quest for gravitational waves (GWs), a primary role among the
possible sources has always been played by those astrophysical events
that are expected to produce GW bursts, such as the gravitational
collapse of stars or the final few orbits and the subsequent
coalescence of a close binary system of neutron stars or black holes.
The search for such transient GW requires the use of a network of
detectors. In fact, the analysis of simultaneous data from more
detectors at different sites allows an efficient rejection of the
spurious outliers, either caused by transient local disturbances
or by the intrinsic noise of the detectors. Resonant GW detectors have
operated for decades in several laboratories around the world, reliably
staying on the air for long periods with high duty cycle
\cite{igec1, igec2, igec3},
mainly looking for burst events. The coming of age of laser
interferometer detectors \cite{ifo}, with much better sensitivity and
bandwidth, has lead to a gradual phasing out of many resonant
detectors.

The ROG Collaboration has built and operated two cryogenic,
resonant-mass detectors, EXPLORER \cite{expl1,expl2, expl3} at CERN
and NAUTILUS \cite{naut1,naut2}  at the INFN Frascati National
Labs(Italy). 
Both detectors have been on the air since the early '90s, performing various 
joint coincidence searches\cite{Explnau1, Explnau2, Explnau3}.
In the period  May 5th 2005 to April 15th 2007 they 
took  part in the IGEC2 network \cite{igec2,igec3} that
collected and exchanged data,
together with the Auriga detector at the INFN Legnaro National Labs
(Italy) \cite{aur} and  with the Allegro detector at LSU (USA)
\cite{all}.
 After that period, Allegro was shut down and data have
been collected by the three surviving antennas, but never analyzed
before.

All these detectors use the same principles of operation. The GW
excites the odd longitudinal modes of the cylindrical bar, which is
cooled to cryogenic temperatures to reduce the thermal noise and is
isolated from seismic and acoustic disturbances. Both Explorer and
Nautilus consist of a large aluminum alloy cylinder (3 m long, 0.6 m
diameter) suspended in vacuum by a cable around its central section
and cooled to about 2 K by means of a superfluid helium bath.
To record  the vibrations of the bar first longitudinal mode,
an auxiliary mechanical resonator tuned to the same frequency
is bolted on one bar end face. This resonator is part of a capacitive
electro-mechanical transducer that produces an electrical a.c. current
that is proportional to the displacement between the secondary
resonator and the bar end face. Such current is then amplified by means
of a dcSQUID superconductive device. Nautilus is also
equipped with a dilution refrigerator that enables operations at 0.1 K,
further reducing the thermal noise. In the period considered, however,
the refrigerator was not operational, in order to maximize the detector
duty cycle. Both detectors are equipped with cosmic ray telescopes, to
veto excitations due to large showers\cite{cosmna1,cosmiciEX,cosmiciNA}.
The two telescopes rely on different technologies (scintillators for
Explorer, streamer tubes for Nautilus) but both provide a monitor of
comparable effectiveness and a continuous check of the antenna
sensitivity\cite{hicosmna,rap}.

At present, while the large interferometers VIRGO and LIGO are
undergoing massive overhauls to upgrade their sensitivity, there
still are two resonant detectors, Nautilus and Auriga,
that continue to operate
in "astro-watch" mode, i.e. as sentinels recording data that could be
analyzed in conjunction with a significant astrophysical trigger, such
as the explosion of a nearby supernova, or any astronomical event
thought to be a possible source of GW.

\begin{figure}
\includegraphics[width=8cm,height=7cm]{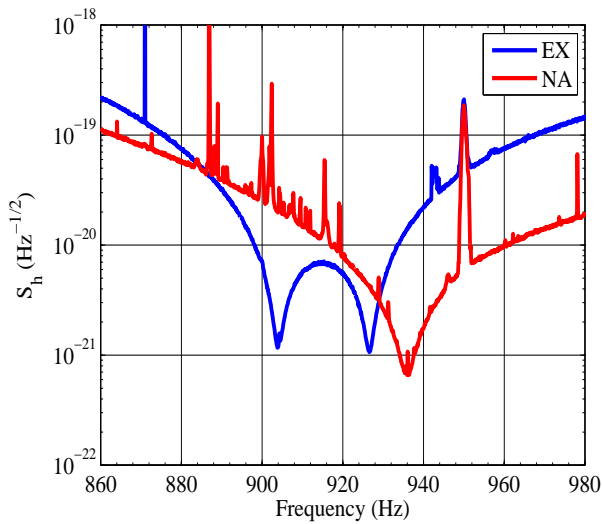}
\caption{Spectral sensitivity curves of EXPLORER and NAUTILUS.
The two bandwidths overlap for a large fraction of the total
sensitive region.}
\label{fig_Sh}
\end{figure}

We report here a study on three years of data from Explorer and
Nautilus,  starting from the end of the IGEC2 network, April 16,2007
and stretching till  June 10, 2010, when Explorer ceased operations.
The spectral sensitivity of the two detectors is shown in
fig.\ref{fig_Sh}.  

The purpose of this paper is to describe a search
for short burst coincident events in the 3 years of data.
The main interest of this analysis lies in several novelties that
were implemented in the data analysis procedure and that are here
detailed; namely, the construction of {\emph Receiver Operating 
Characteristics} for the 2-detector
observatory, in addition to the ones for each antenna, the optimization
of the threshold pairs in order to maximize the detection efficiency,
given an a-priori choice of the background of accidental coincidences
and an optimized procedure for the calculations of upper limits on
the rate of GW burst. The search
was carried out keeping a low  level of accidental coincidences, that
we set at 0.1 over the entire period. Along with this search, we also
performed a calculation of the upper limit (UL) on the rate of
delta-like GW pulses impinging on the Earth. The method here described
presents relevant improvements with respect to previously published
searches performed with resonant detectors: we have used software
injections of known signals to measure the efficiency of detection
for each antenna and for the combined observatory. Based on the
efficiencies so evaluated, and on the measured rate of accidentals, the
analysis parameters were separately optimized for the coincidence 
search and for the UL evaluation.

Throughout this paper, we shall call "events" or "outliers" those
data points in the filtered data stream
that are larger than a given threshold: these points are selected by an
automatic event finder procedure and constitute the database for our
analysis. 

The paper is organized as follows: in sect.\ref{sect_data} we
describe the data collected in these 3 years, the filtering procedure,
the criteria chosen to segment the total observation period in 5
subperiods and the vetos applied to both data and events. In
sect.\ref{sect_dect} the procedures of software injection and
time-delayed coincidences are detailed. The detectors are characterized
in terms of efficiency and accidentals in the various time segments and
the ROCs are generated, both for the individual antennas and for the
combined observatory; on the basis of these parameters, in 
sect.\ref{sect_coinci} the thresholds for the "true" (zero-delay, on
time) coincidence search are chosen and the search is performed.
Finally, in sect.\ref{sect_UL}, we describe the procedure used to
compute the upper limit for the flux of GW radiation.
This procedure, quite different from those used in the past, is
optimized in each of the subperiods and for each of the amplitudes of
GW signals considered.

Some final considerations conclude the paper.

\section{The data}\label{sect_data}

Data are collected by the two detectors with almost identical hardware
and software. The output of the SQUID amplifier is conditioned by band
pass filtering and by an anti-aliasing low-pass filter, then sampled at
5 kHz and stored on disk.
Sampling  is triggered by a GPS disciplined rubidium oscillator, also
providing the time stamp for the acquired data.

The data are processed off-line, applying adaptive, frequency domain
filters.
We first "whiten" the data,  i.e. remove the effect of the
detector transfer function, producing the so called $reconstructed ~h$. 
A filter matched to delta (or very short) excitations is then applied
to this stream. 
We shall call $h_{\delta}(t)$ the output of this process, to remind
that the time series so produced
does not represent a generic wave amplitude $h$, but is the
best effort to detect an impulsive excitation.
The  noise characteristics
estimate is updated averaging $h^2_{\delta}(t)$ over 10 minutes periods.
The filters used in the above procedure
are computed using a model for the detector, fitted
with the measured values of frequencies and decay times of the system
resonances, and the experimental noise spectrum.
Both the model and the signal response were validated by
hardware injections of known signals: the filtered output matched the
expected value to better than $10\%$.

As usually done for resonant detectors \cite{igec2},
the filtered output $h_{\delta}(t)$ is normalized assuming that
the excitation lasts $\simeq 1 ~ms$ and has a bandwidth of
$\simeq 1 ~kHz$ centered and flat in the region of the detectors
sensitivity.
If these assumptions are fulfilled, the peak value of $h_{\delta}(t)$
gives the $h_{rss}$ of the input signal.

The filter is designed and optimized for delta-like signals, but it
works equally well \cite{IGEC3} for a wider class of short
bursts, like e.g. damped sinusoids with decay time $\tau < 5$ ms.  
Typical GW signals of this kind and their possible astrophysical sources
have been discussed and exposed e.g. in \cite{IGEC3} and
references therein.

Although the detectors produce quite stationary data, their
characteristics did change a few times over such a long observation
period: in some instances, these differences were due to actual changes
in hardware (e.g. substitution of a preamplifier), other times to some
non identified factors.  We found it useful, to the purpose of the
study detailed below, to segment the analysis in different periods
where both detectors had noise behavior (average noise energy)
\footnote{Traditionally, the GW resonant detectors community indicates
the antenna sensitivity, i.e. the long term quadratic average of the
filtered output $h_{\delta}(t)$, in terms of a "detection noise
temperature"  $T_{eff}$. These two quantities are related, for our
detectors, by $<h^2_{\delta}>  = 6.35\E{-35} ~T_{eff}$. In these units,
the average noise levels of table I spanned between 0.9 and 4 mK.}
consistently stable, within the statistical fluctuations. This allows
us to better optimize the search in each period.
Consequently, we identified 5 time stretches (see table I),
that roughly coincide with solar years, and ran separate optimized
analysis on each subperiod. Stretch $\#2$, the end of 2007, covers a
short period, when Nautilus operations were badly disturbed: we shall
show that the adopted procedures automatically minimize the
contributions from bad periods both for the coincidence search and
for the upper limit evaluation.

Before we start describing the procedure two comments are in order : \\
1) when segmenting the data in subperiods to be treated separately, or
to run separate optimizations of the search parameters,  we must ensure
that each subperiod be long enough to provide a sufficient statistics,
and in particular to avoid that any particular outlier or
temporary noise affect in a sizable way the final choices. We
found that a few days of data is somehow the minimum duration
for this purpose.\\
2) if the procedure is properly devised, 
the addition of any information or data set, however
poor its quality with respect to the rest of the data, should
not reduce the quality  of the total result. On the contrary,
a correct way of putting together all the information can only
produce a better result.

\begin{table}[h!]
\centering
\begin{tabular}{||l||c|c|c|c||}
\hline\hline
 % & & & &\\[-3mm]
  Subperiod & & Days of  & \multicolumn{2}{|c||}
{$\sqrt{<h^2_{\delta}>} \E{-19}$} \\
   \hskip 1cm $T_i$ & Begin - End     &good data& Explorer &Nautilus \\
\hline\hline
& & & & \\[-4mm]
 \#1 : 2007 A & Apr. 16 - Dec. 5 & 162 	&4.57  & 3.47\\
 \#2 : 2007 B & Dec. 6 - Dec. 31 & 12  	& 4.21 & 5.04\\
 \#3 : 2008   &Jan. 1  - Dec. 31 & 232 &  4.43& 4.36\\
 \#4 : 2009   &Jan. 1 -  Dec. 31 & 242 & 3.82 &2.52 \\
 \#5 : 2010   & Jan. 1  - Jun. 10  & 113 &3.98  & 2.39\\[-4mm]
 & & & & \\
\hline\hline
\end{tabular}
\label{tab_per}
\caption{The time stretches $T_i  (i=1...5)$ used to run different
optimization in our analysis.
$\sqrt{<h^2_{\delta}(t)>}$ is the long term average of the amplitude
noise}
\end{table}

\subsection{Data selection}
The data  were selected with different cuts, applied both to the data
stream and to the list of outliers. All criteria, studied
in the past and  in use for several years, were a priori
chosen and blindly applied.
The vetos that cause elimination of entire periods of data
stream include:
\begin{itemize}
\item when acquisition flags or operator's notes are recorded,
indicating bad or suspected periods (e.g., cryogenic refills, activity
around the detector....)
\item when  the noise of the filtered data, averaged over 10 minutes, 
rises above a given value (about 5 times the long term average)
\item  when the reference tone, a monochromatic signal monitoring the
gain of the electronic chain, falls outside a given range
\item when an excessive amount of wide-band noise is present.
Wide-band noise, usually of electronics origin, is monitored on two
frequency bands, above and below the useful bandwidth of the detector.
\item  when auxiliary channels exhibit  mean values above predetermined
levels. Auxiliary (or veto) channels include seismic monitors, SQUID
locking working point, nitrogen (on Explorer) and helium flow and more.
\end{itemize}

These cuts reduce the amount of available data for the coincidence
analysis to 761 days, i.e. two thirds of the 1152 days of total
observation period. The main contribution to these cuts is due to
operations of cryogenic maintenance (liquid helium refills) that we
chose to perform in different times on the two detectors, so that at
least one were always operational.

On these data, an automatic event finder procedure selects the
"outliers". 
All data points remaining above a chosen threshold are grouped in one
$event$. An event can extend over more than one group if the signal
falls below threshold for a time shorter than the $dead ~ time$, set
to 1 s.  
For the class of short signals discussed above, the shape of the event
is mostly due to the antenna response function (see e.g.
fig.\ref{fig_cosm}). Each event is then characterized by the time and
amplitude of the largest sample.
%The relation between the peak output value $h_{max}$ and the input
%signal {\it root sum square amplitude} $h_{rss}$ is simply assumed 
%to be $h_{rss} = h_{max} \sqrt{\tau}$,
%where we chose, as usually done for resonant detectors \cite{ROG2006},
%$\tau=1$ ms.
Further characteristic are recorded for each event, such as: starting
time, total time length, integrated amplitude of the samples above
threshold, average noise before the event.    
We set for this selection a threshold at critical ratio CR=5 with
respect to the average noise level, continuously updated.

A further selection was then applied to the outliers, in order
to implement other cuts:
\begin{itemize}
\item  an event should remain above threshold for a time
consistent with its amplitude (the decay time for the filtered data
is the inverse of the detector bandwidth).
\item cosmic ray showers are known to produce short bursts of
excitation in the antennas.  
The events must not be in coincidence with a shower, as
recorded by detectors installed above and below both antennas
\cite{cosmiciEX,cosmiciNA}. 
\end{itemize}

These two selections veto a very small fraction of the
events, usually less than 0.1\%.

\section{Detectors characterization} \label{sect_dect}

In order to perform a "fair" search for coincidence, all "human
handles", i.e. adjustable parameters, must be a priori set before
starting the search.
To this purpose, we have applied a very large number of software
injected events to determine the efficiency of both detectors to
short bursts of GW.
Likewise, the background of accidental coincidences is determined
via a large number of time shifts. For a given level of accidentals,
a priori set, the efficiency is then maximized with a proper choice
of the thresholds. Only at this point, we can "open the box", i.e.
look at the zero delay coincidences and assess its significance.
The following subsections give some details about this characterization
procedure.

\begin{table}
\begin{tabular}{||l||c|c|c||}
\hline\hline
 ~~Subperiod & Duration &  $N_{acc}$~ &  $\overline{N}_{acc}$\\
    & $(day)$ & $in ~10^4 ~shifts$ & $/(day\cdot shift)$ \\
\hline \hline
 & & & \\[-3mm]
 \#1 : 2007 A & 162 & 5,635,671  & 3.48 \\
 \#2 : 2007 B & 12  & 640,143    & 5.33 \\
 \#3 : 2008   & 232 & 6,667,062  & 2.87 \\
 \#4 : 2009   & 242 & 3,061,314  & 1.26 \\
 \#5 : 2010   & 113 & 1,042,858  & 0.92 \\[-4mm]
& & &\\
\hline\hline
\end{tabular}
\label{tab_acc}
\caption{Accidental (time-shifted) coincidences ($N_{acc}$) in the
5 subperiods analyzed, obtained with the lowest threshold,
$h_{rss}=3.56 \E{-20} s^{1/2}$, used in this analysis.}
\end{table}

\subsection{Accidentals}\label{sect_acc}

The evaluation of the expected background of accidental coincidence
was performed with the usual method of the time shifts.
The lists of events, one for each detector, extracted from
the data as described in sect.\ref{sect_data}, were compared after
shifting the time stamp of one of them. The event times of one
detector were delayed, with respect to the other ones, 10,000 times
in steps of 1.5 seconds (i.e. between $\pm 7,500 $s, excluding the
zero time shift). This value is larger than the dead time \cite{IGEC3}
inserted by the event finder.
The search of coincidences in each of the 10,000 cases, performed with a
time window of 15 ms as discussed in sect.\ref{sect_eff}, produces the
data base of unphysical coincidences from which we learned the
background characteristics. Tab.II summarizes the results.

\begin{figure}
\includegraphics[width=7cm]{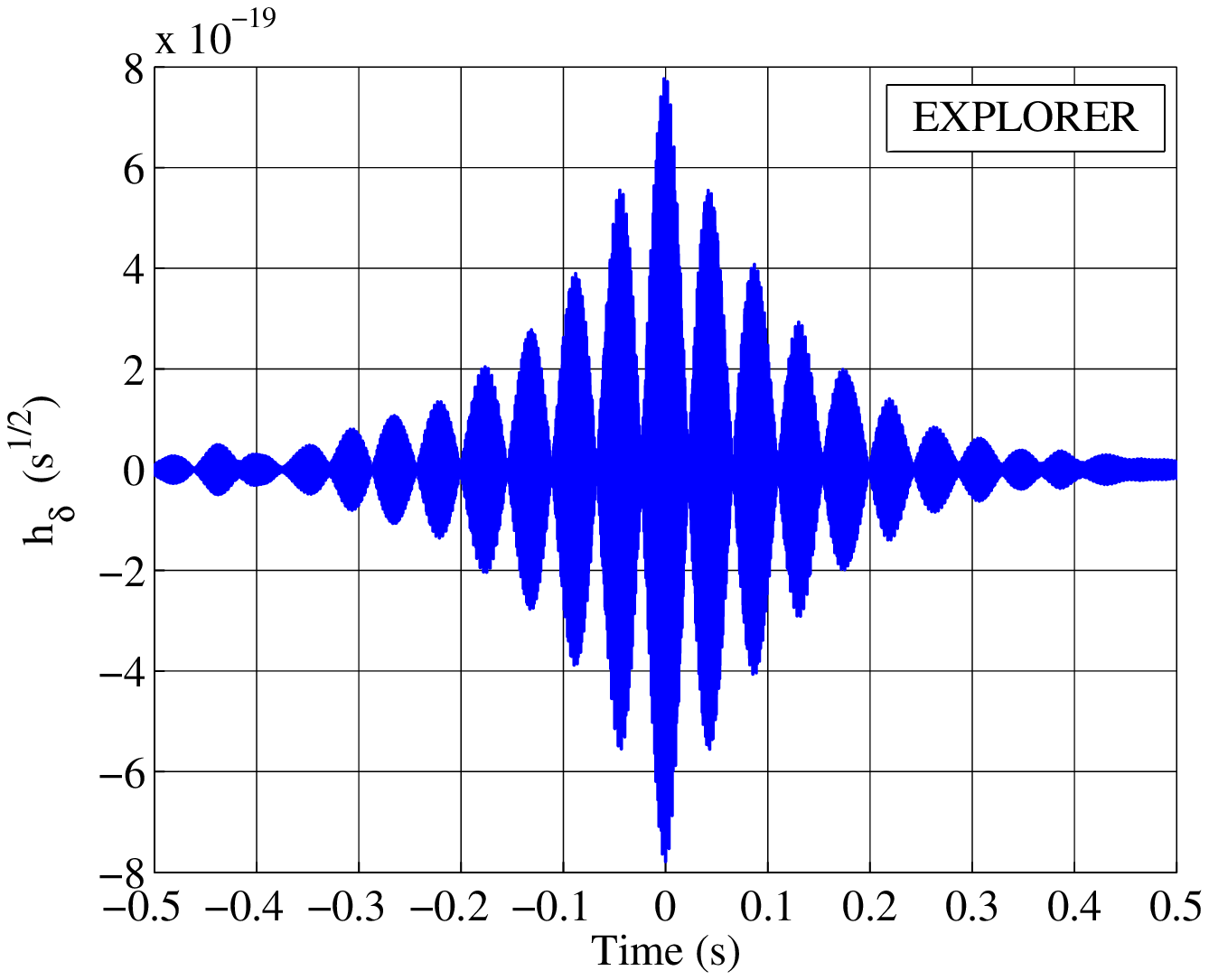}
\includegraphics[width=7cm]{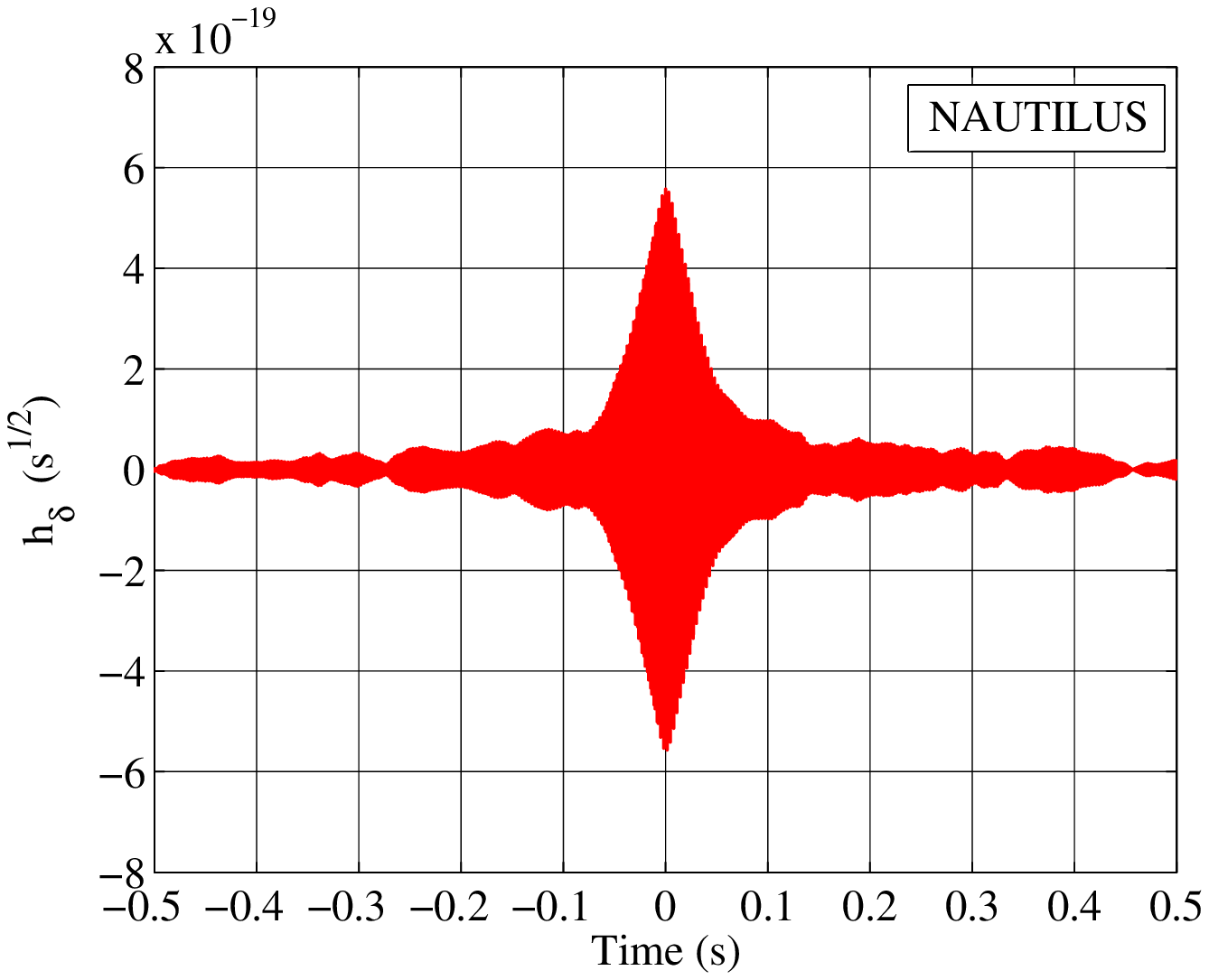}
\caption{Excitation of the detectors, as seen in the filtered data,
 due to two large cosmic-ray
showers: an event of Explorer 2008 and an event of Nautilus
2007.
Only the maximum value of these signals can be interpreted in
terms of the assumed $h_{rss}$ excitation; the shape of the pulse is
due to the antenna response. The two plots are so dissimilar because of
the differences in the bandwidths (see fig.\ref{fig_Sh}): while Nautilus
has its sensitivity around one main frequency, Explorer is most
sensitive on two frequencies and therefore its time-domain response
exhibits beats.}
\label{fig_cosm}
\end{figure}

\subsection{Software injections} \label{sect_inj}
Large sets of software injections were performed in order to determine
the efficiency of the detectors to delta-like signals of different
amplitudes. As mentioned above, the extensive cosmic ray showers excite
the bars, closely approaching the effect of a short GW burst.
We took advantage of this feature and used real signals, observed in
coincidence with some particularly intense cosmic ray shower, as the
prototype signal to be used for software injections (see
fig.\ref{fig_cosm} for an example of the signals applied).
These signals, actually oversampled at 50 kHz, were scaled to the
appropriate values of amplitude and added to the filtered data of
each detector. 
This technique is much faster than that generally used, where 
one first generates the h-reconstructed data stream, then adds the
injections to this stream and finally re-filters and searches for the
events. We validated our method by applying both techniques
to a one-day sample of data, finding a very good agreement.
The times of the injections were pseudo random, because we
avoided injections too close to the beginning or end of each period
of good data, and required a minimum distance of 10 seconds between
two adjacent injections.
Moreover, we added a delay, randomly chosen in the $[-2.3, 2.3]$ms
interval to the injection time of Explorer, to simulate the time of
flight of a possible GW signal of unknown direction. 

We injected signals of 10 different amplitudes, in the range of
$h_{rss} ~[7.97 \div 12.6] \cdot 10^{-20} s^{1/2}$, at a rate of about
90 injections per day.

\subsection{Efficiency} \label{sect_eff}
\begin{figure}
\includegraphics[width=7cm]{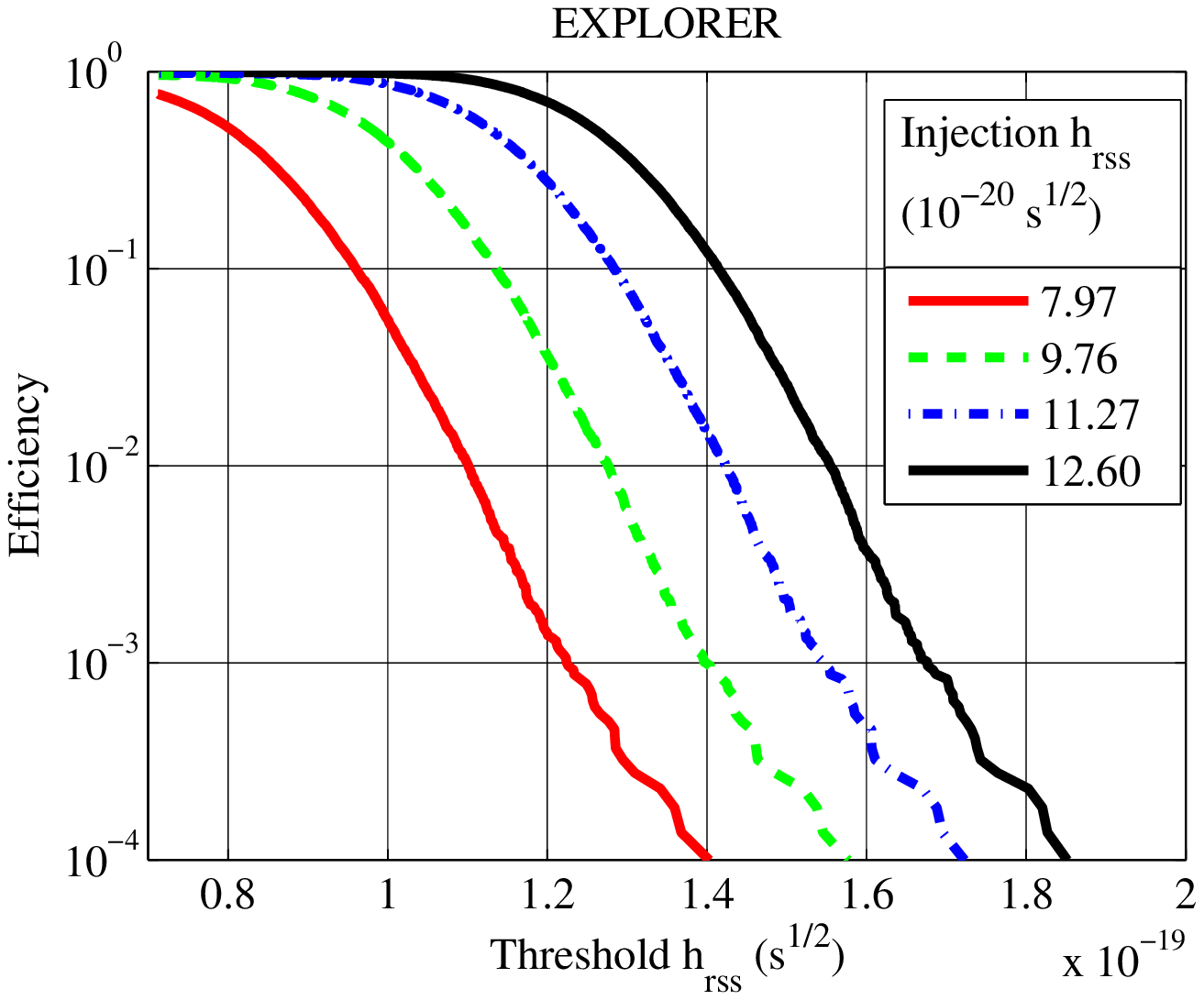}
\includegraphics[width=7cm]{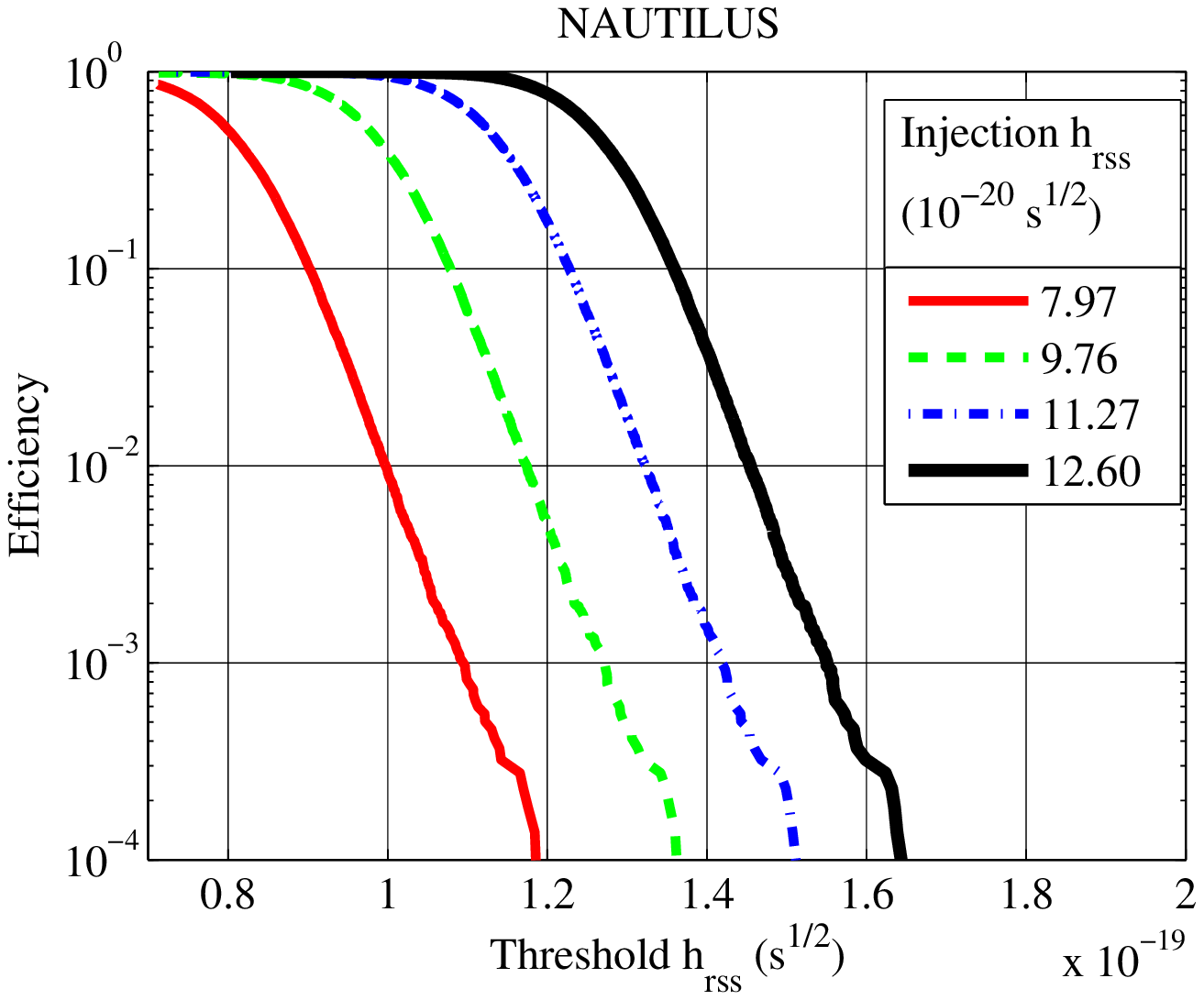}
\caption{Efficiencies of Explorer and Nautilus in 2009.
The four lines refer to injections with 
$h_{rss}=(7.97,~9.76,~11.27,~12.6)\E{-20} s^{1/2}$. }
\label{fig_effic}
\end{figure}

The usual event finder routine
was then  applied to the data containing the injected signals. 
For each sub-period and for each level of injected signal, efficiency
charts were produced, displaying the percentage of detected signals
with amplitude exceeding any given value.
Fig.\ref{fig_effic} is a sample of such charts, showing the
efficiencies in the subperiod 2009 for each antenna.  

The injections also allow us to determine the time response of
the detectors, and guided us in choosing the best coincidence time
window to be applied. We found that a coincidence window of
$\pm 15$ ms assures an efficiency very close to 1 for delta-like
signals, even at the lowest injected amplitude: indeed, the 
measured probability of missing a coincident event with the chosen
window of 15 ms is less than $ 1\E{-4}$.
Besides, the chosen window is sufficiently wide to also accommodate,
without significant losses of efficiency, other classes of signals
\cite{IGEC3} for which the detectors time response
might be different.

\subsection{Receiver Operating Characteristics}\label{ROC}

Efficiency and accidentals vs threshold amplitude completely
characterize a detector.
These two classes of information can be summarized in the
\emph{Receiver Operating Characteristics} or ROC. 

 It is worthwhile, in view of what follows, to briefly recall the
procedure to generate a ROC:
for each  injected amplitude we sweep the threshold amplitude and we
look up both the efficiency and the rate (or the total number) of
accidentals. 
By eliminating the threshold value between 
these two relations, we derive a curve [efficiency vs event rate],
that constitutes the ROC for that given signal amplitude.

\begin{figure*}
\includegraphics[width=7.5cm,height=5cm]{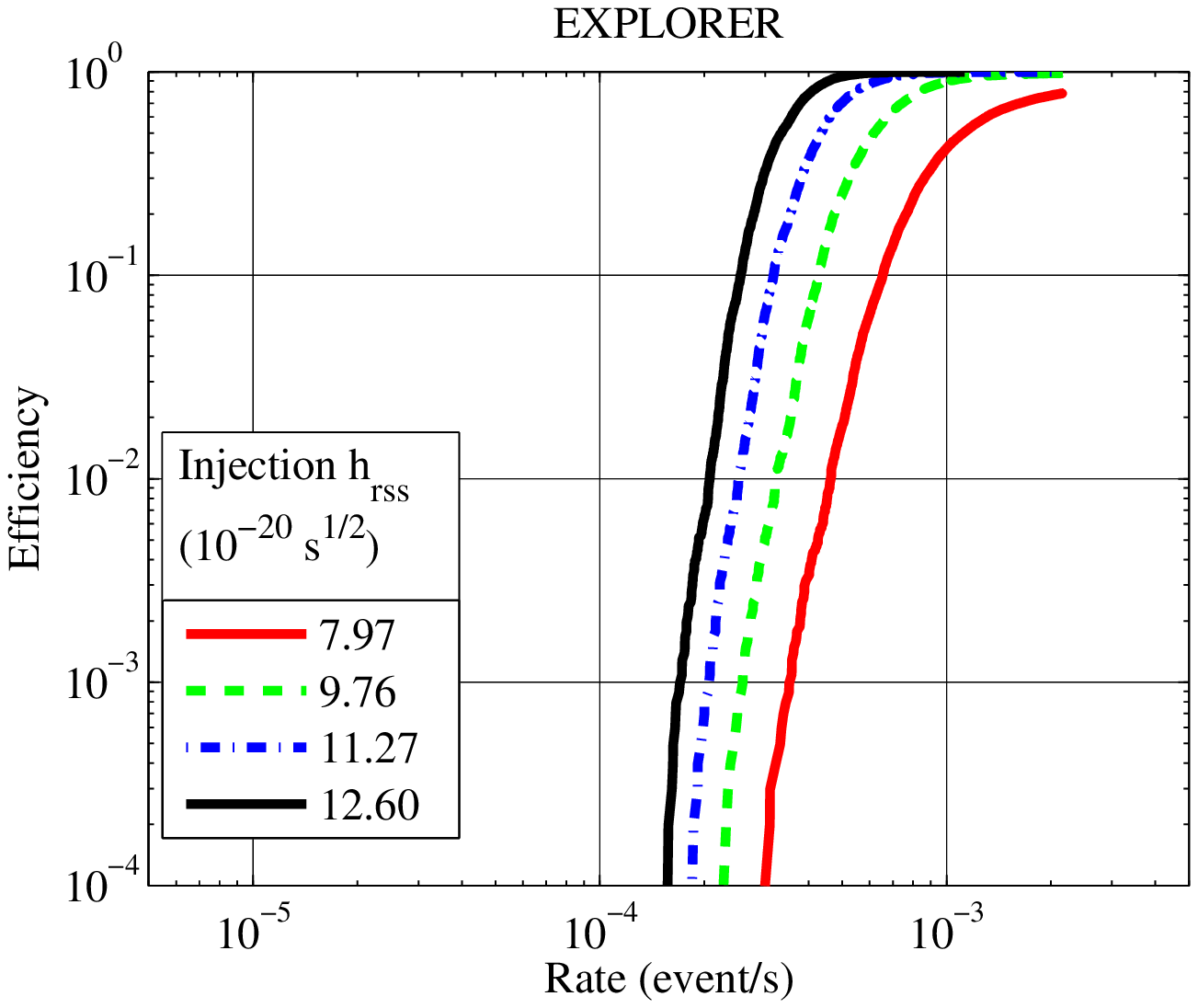}
\includegraphics[width=7.5cm,height=5cm]{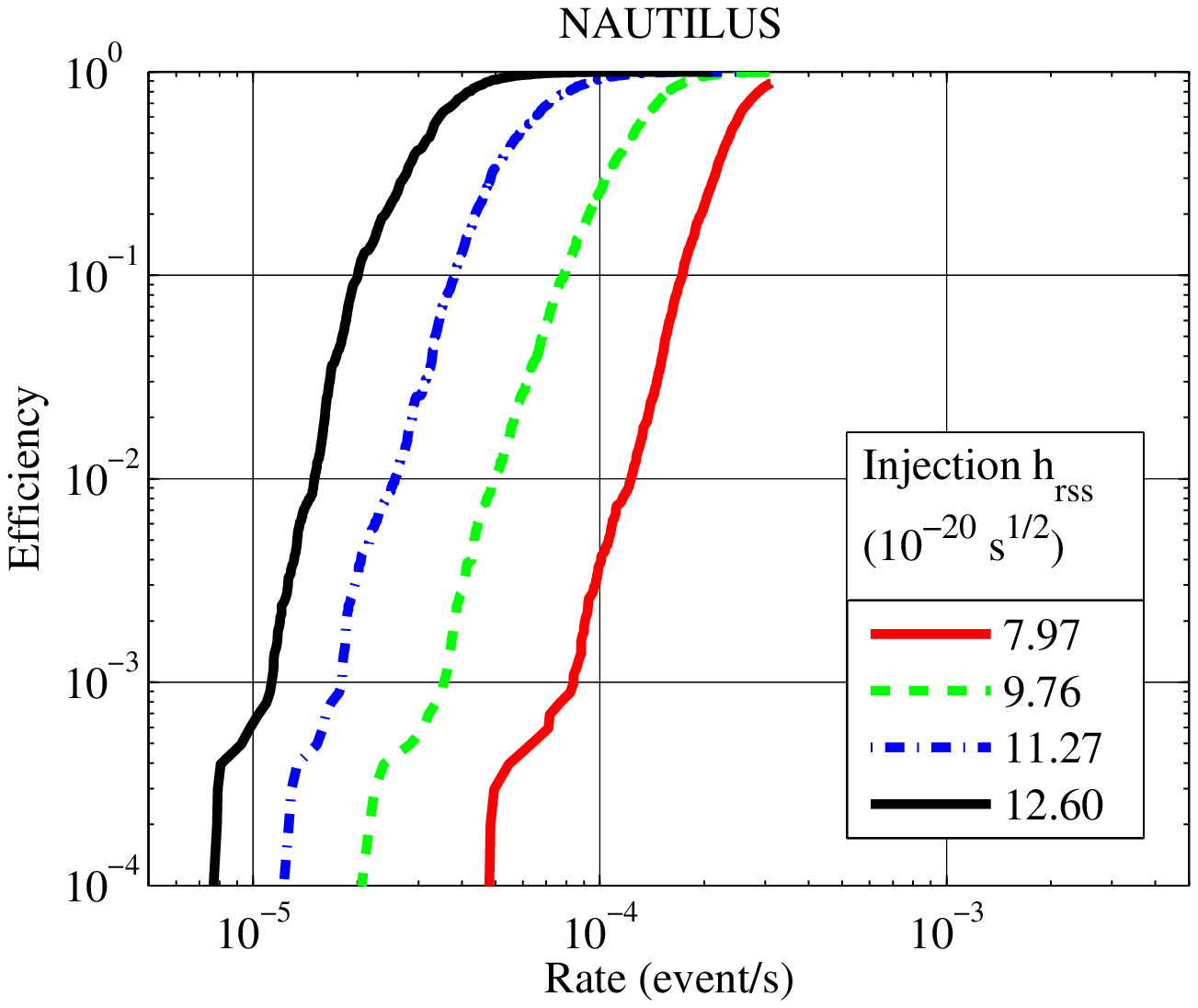}
\caption{ROCs for Explorer and Nautilus in year 2010.}
\label{fig_roc_exna}
\includegraphics[width=7.5cm,height=5cm]{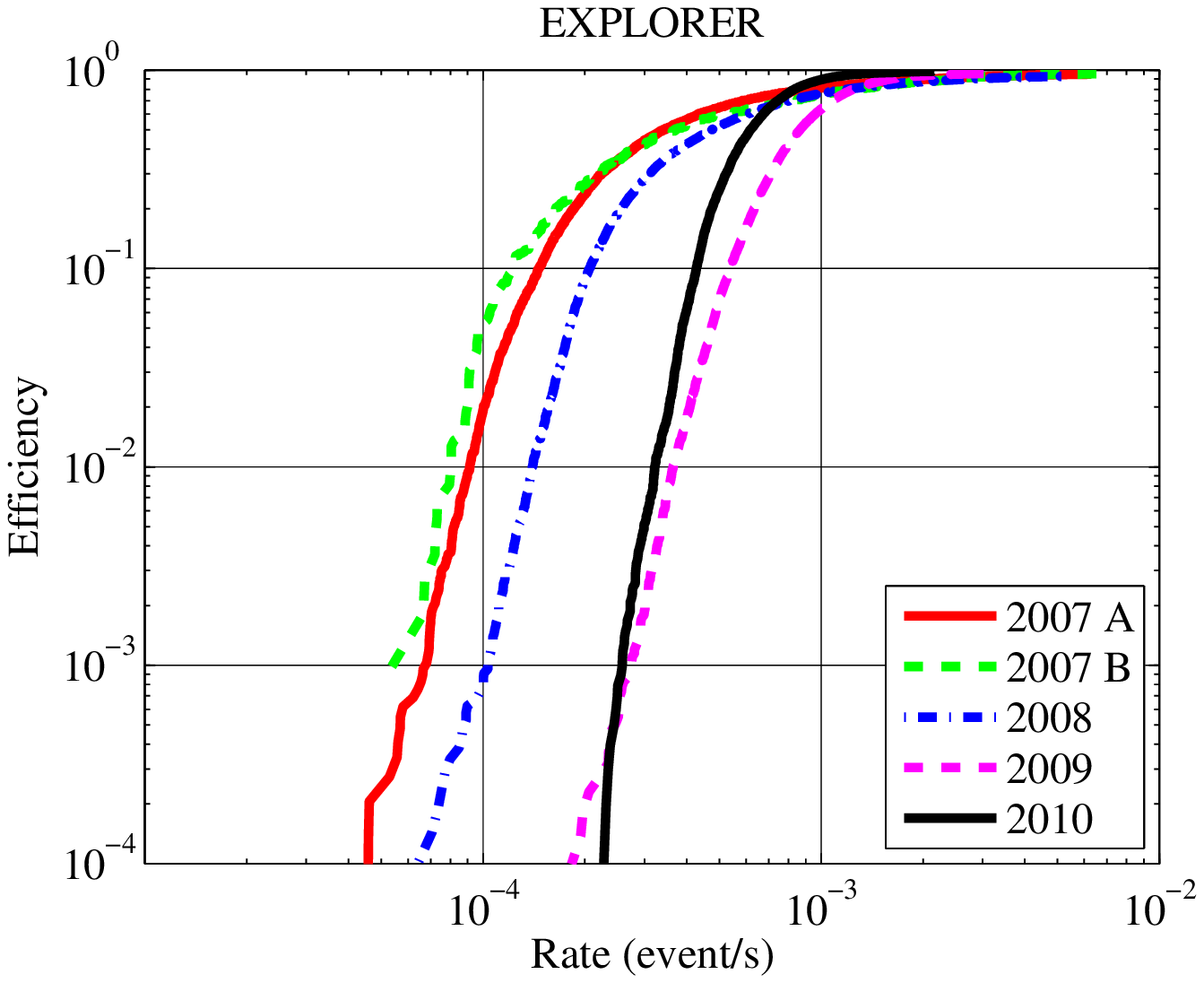}
\includegraphics[width=7.5cm,height=5cm]{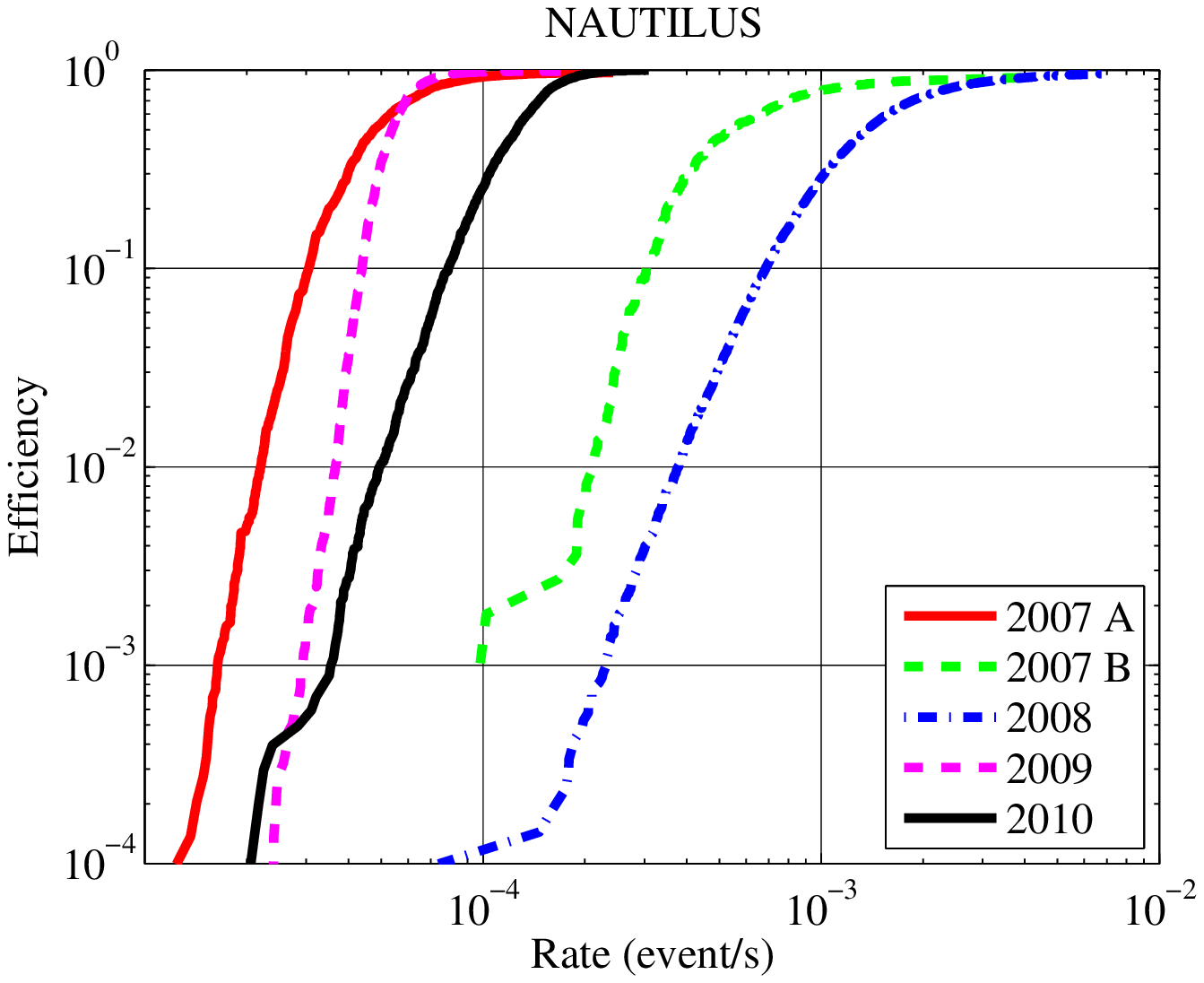}
\caption{ROCs for Explorer and Nautilus, at an injected amplitude
$h_{rss}=9.76 \E{-20} s^{1/2}$ in the five subperiods. }
\label{fig_roc_5subper}
\includegraphics[width=7.5cm,height=5cm]{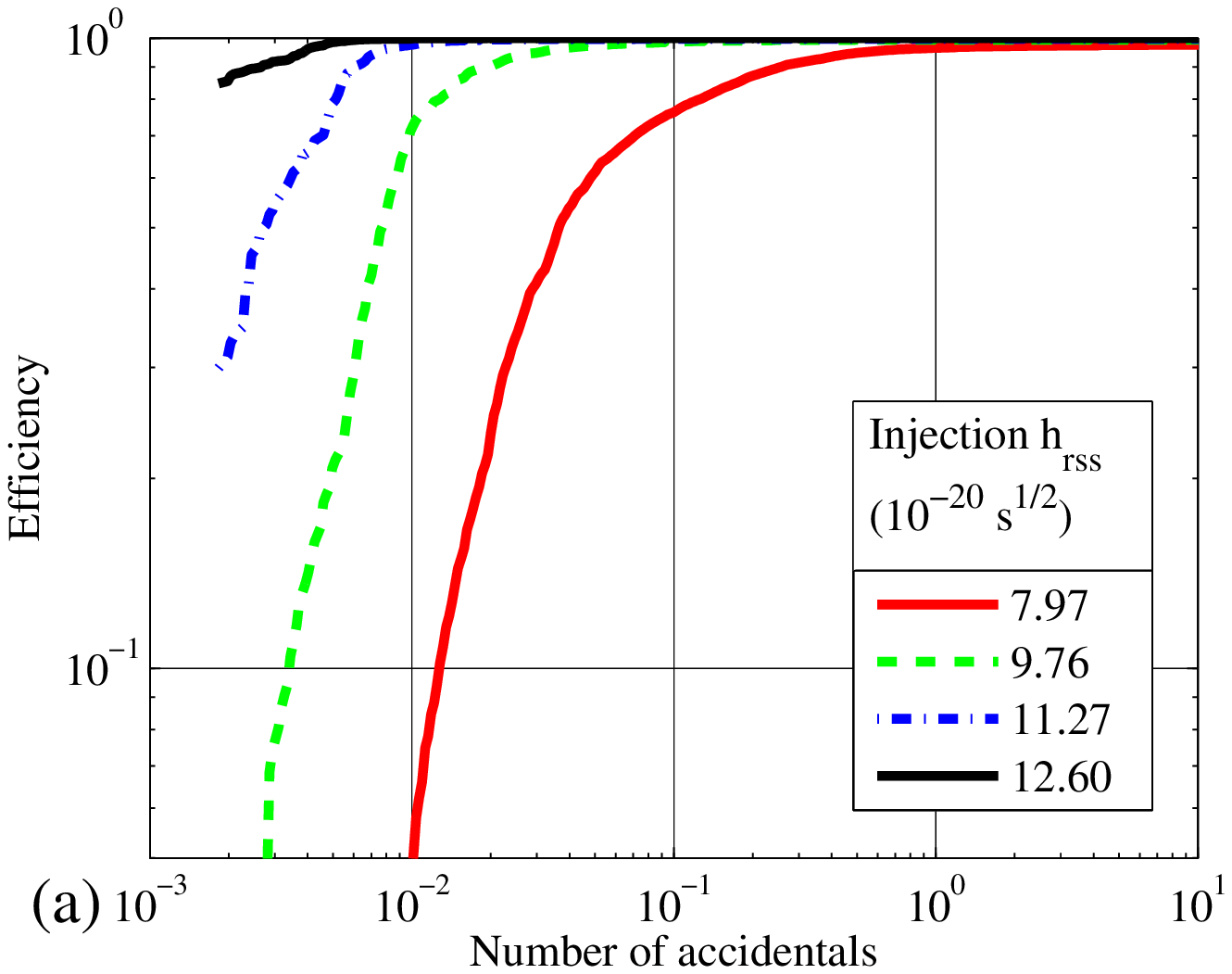}
\includegraphics[width=7.5cm,height=5cm]{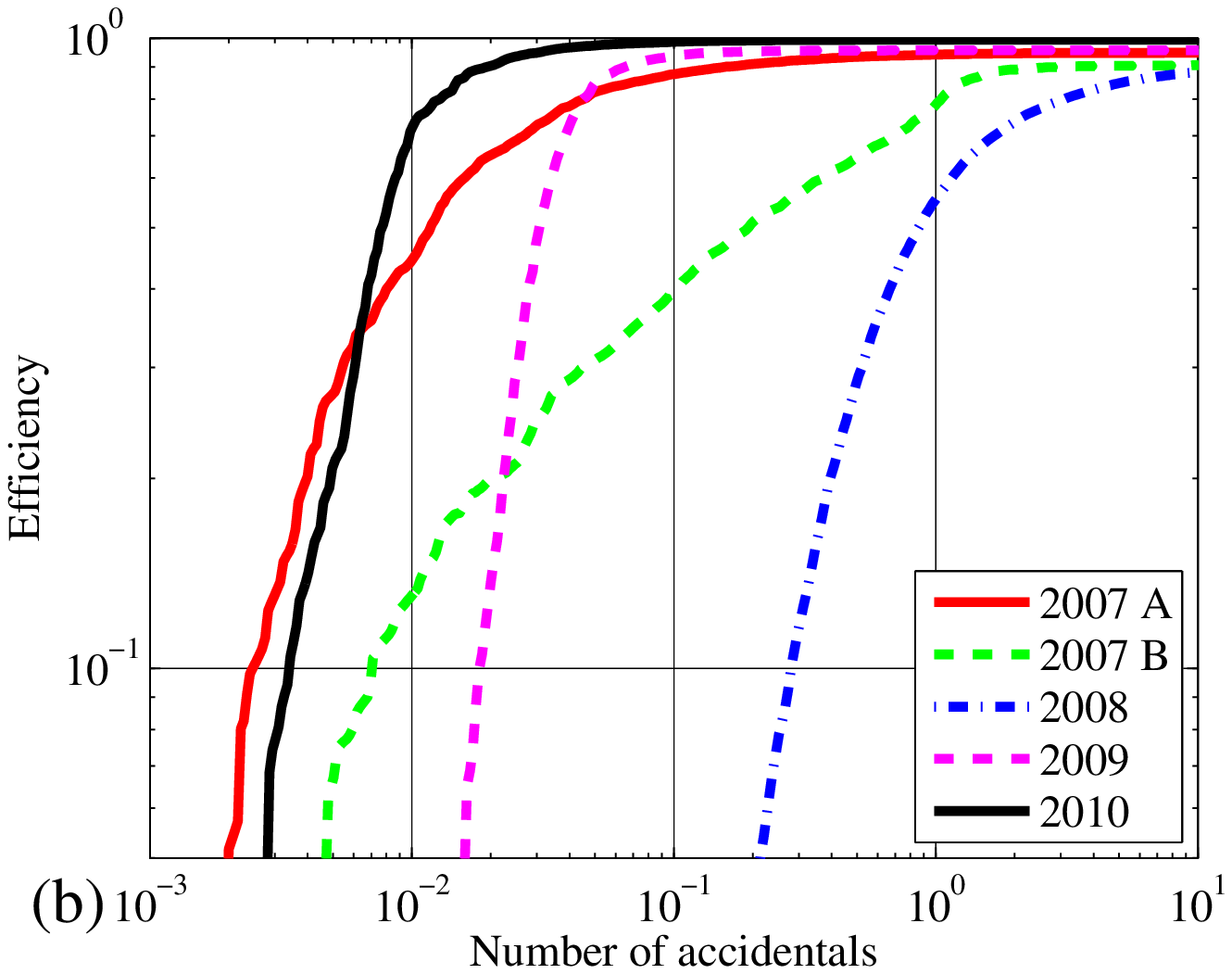}
\caption{ROCs for the combined observatory (Explorer + Nautilus).
Left: in 2010, at the four injection amplitudes considered in
fig.\ref{fig_roc_exna}; right: at an injected amplitude
$h_{rss}=9.76 \E{-20} s^{1/2}$ in the five subperiods. }
\label{fig_roc_coi}
\end{figure*}

In fig.\ref{fig_roc_exna} we show an example of ROCs, for both
Explorer and Nautilus, relative to year 2010.

It is to be remarked that, despite the fact that the detector hardware
was virtually unmodified in all subperiods, the ROCs do vary,
especially for Nautilus, from one subperiod to another (see
fig.\ref{fig_roc_5subper}).

As we are interested in the operation of both antennas as one detector,
we can extend the concept of ROC to a coincidence search.
In this case,  in order to vary efficiency
or accidental rate, we can act on either threshold, so that there
exists an infinity of threshold pairs that can provide the same
characteristics: we could have therefore infinite ROC curves for the
same signal amplitude.
However, keeping in mind that our aim is to maximize the efficiency
of detection for a given accidental rate, we can focus our search on
finding the pair of thresholds that gives the best  efficiency for
each value of accidentals.
The ROCs for the Explorer + Nautilus compound observatory are therefore
obtained with the following procedure:
\begin{itemize}
\item we choose a set of M threshold values and we sift through our
data with a matrix  of M*M  thresholds:  both in the list of events
found with the injections where we demand triple coincidences ($t_{EX},
t_{NA}$ and $t_{inj}$), and in the set of shifted coincidences, for
the accidentals.
\item  in this way we create a M*M matrix with values of efficiency
and accidentals for each threshold pair.
\item  for each of N chosen values of accidentals, we search the
matrix for those intervals that contain that value of accidentals.
We interpolate in those intervals to find the value of thresholds and
efficiency.
\item  finally, we compare these values and choose that with the
largest efficiency.
\end{itemize}
In our search, we used N=M= 100 and repeated the procedure for the 5
subperiods and for each of the 10 values of injected amplitudes.

Some of the ROCs for the observatory, obtained with this procedure,
are shown in fig.\ref{fig_roc_coi}.
Each point of  these curves represents the threshold pair that
produces the desired value of accidentals with the best possible
efficiency. This procedure yields approximated values for the data
points, as they are obtained via interpolation: for this reason, the
search was later refined around the selected threshold values.

\section{Coincidence search}\label{sect_coinci}

In order to perform the true-time (on-source) search, we must decide
upon a unique set of 5*2 thresholds to be applied to the 2 detectors
in each of the 5 subperiods. This set must provide the desired number
of total accidentals (0.1) while achieving the maximum possible
efficiency for GW signals.

The ROCs for the coincidences, previously determined, specify, in each
subperiod and for each injection amplitude, what are the thresholds
capable of obtaining a given value of accidentals with the maximum
possible efficiency. Next step is to find how to distribute the total
number of accidentals between the different periods in order to
maximize the total efficiency defined as
\begin{equation}
\overline{\varepsilon} = \sum_{i=1}^{5} \frac{\varepsilon_i T_i}{T}
\end{equation}
where $\varepsilon_i, T_i$ are the efficiency and duration of the
subperiods and $T =  \sum_i T_i$ is the total observation time.

\begin{figure}
\includegraphics[width=8cm,height=7cm]{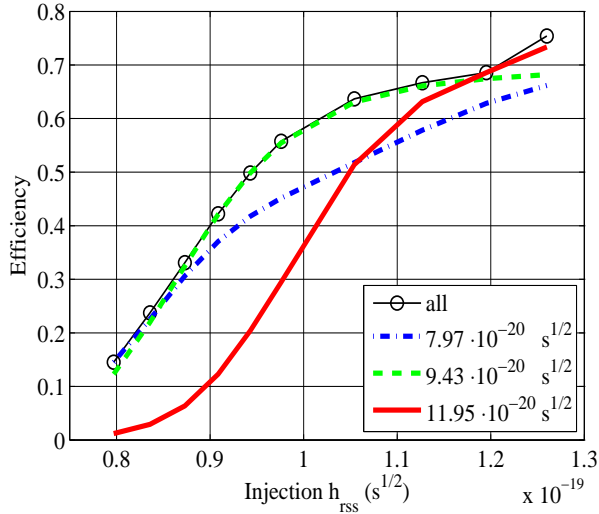}
\caption{Maximum efficiency $\overline{\varepsilon}$ achievable with a
background set at 0.1 accidentals in the entire observation time. The
top line shows the efficiency with thresholds optimized at each
abscissa point (injected amplitude), while the other curves are
produced with different choices of a unique set of thresholds.}
\label{fig_acc01_m}
\end{figure}

We remark that this procedure pins down a different set of thresholds
for each considered injection amplitude. 
In fig.\ref{fig_acc01_m}, the data points show
the results of this optimization: each data point is obtained with its
own optimized set of 10 thresholds; we call this curve "composite
efficiencies".
However, as the coincidence search has to be performed only once, we
need a strategy to select a unique set of thresholds.
Fig.\ref{fig_acc01_m} also shows three curves describing the efficiency
at all amplitudes for three selected sets of thresholds, namely those
optimized for $(7.97,~ 9.43,~11.96)\E{-20} s^{1/2}$. We make here no
assumption on the amplitude distribution of the GW signals we search
for; therefore we selected a threshold set that best approaches the
curve of "composite efficiencies" at all amplitudes, and in particular
at the smaller ones (that are, in such a search, the most probable).

Clearly, the set of thresholds chosen for
$h_{rss}=9.43 \E{-20} s^{1/2}$ (values detailed in table III) is
the one that performs best and was therefore selected. 	

\begin{table}
\centering
\begin{tabular}{||l||c|c||}
\hline\hline
 & \multicolumn{2}{|c||}{} \\[-3mm]
  ~~~Period & \multicolumn{2}{|c||}{$h_{rss}^{thr} [\E{-20} s^{1/2}]$}  \\
 & Expl & Naut \\
\hline\hline
 & &  \\[-4mm]
 \#1 : 2007 A & ~~~9.00~~ & 8.52 \\
 \#2 : 2007 B & 12.3 & 9.60 \\
 \#3 : 2008   & 10.8 & 19.3 \\
 \#4 : 2009   & 8.12 & 8.17 \\
 \#5 : 2010   & 8.19 & 8.29 \\[-4mm]
 & &  \\
\hline\hline
\end{tabular}
\label{tab_val}
\caption{The set of thresholds chosen for the final coincidence search:
these values are optimal for an injected signal with
$h_{rss}=9.43 \E{-20} s^{1/2}$, but well approach the best possible
efficiency at all signal amplitudes, as shown in fig.\ref{fig_acc01_m}.}
\end{table}

\begin{table}
\centering
\begin{tabular}{||c||c|c|c|c|c||}
\hline\hline
 & \multicolumn{5}{|c||}{Sub Period} \\
 & 2007 A & 2007 B & 2008 & 2009 & 2010 \\
 & & & & & \\[-4mm]
\hline\hline
 & & & & & \\[-4mm]
 Accidentals & 0.0206 & 0.0 & 0.0037 & 0.0585 & 0.0172 \\
\hline\hline
 &  \multicolumn{5}{|c||}{} \\[-4mm]
   Injected $h_{rss}$ & \multicolumn{5}{|c||}{Efficiency} \\
   $ [s^{1/2}]\cdot10^{-20}$ & \multicolumn{5}{|c||}{} \\
\hline %\hline
 & & & & & \\[-4mm]
  7.97 & 0.0935 & 0.0    & 0.0 & 0.2066 & 0.1520 \\
  8.36 & 0.1826 & 0.0018 & 0.0 & 0.3787 & 0.3162 \\
  8.73 & 0.2936 & 0.0018 & 0.0 & 0.5542 & 0.5137 \\
  9.09 & 0.4167 & 0.0036 & 0.0 & 0.7063 & 0.6856 \\
  9.43 & 0.5438 & 0.0100 & 0.0 & 0.8143 & 0.8081 \\
  9.76 & 0.6528 & 0.0190 & 0.0 & 0.8840 & 0.8869 \\
 10.54 & 0.8476 & 0.0805 & 0.0 & 0.9533 & 0.9712 \\
 11.27 & 0.9358 & 0.2081 & 0.0 & 0.9758 & 0.9920 \\
 11.95 & 0.9709 & 0.4009 & 0.0 & 0.9843 & 0.9963 \\
 12.60 & 0.9834 & 0.6091 & 0.0 & 0.9890 & 0.9980 \\[-4mm]
 & & & & & \\
\hline\hline
\end{tabular}
\label{tab_varie}
\caption{Accidentals and efficiencies at various amplitudes with the
chosen set of thresholds (see tab.III). Note that 2007B gives no
contribution to the accidental background, while for 2008 the
efficiency results set to zero at all amplitudes. }
\end{table}

Table \ref{tab_varie} shows how the overall background was distributed
and how the efficiency of detection at several signal amplitudes
changed over the 5 subperiods of the search. We note that the
optimization procedure automatically weights the subperiods according
to the data quality, virtually "turning off", without any manual
adjustment,  the noisiest periods, i.e. 2007B and 2008: where we have
a noisier detector, there we get little or no contribution to the
coincidence search.

When we finally applied this set of thresholds to the on-time data, no
coincident events were found, thus returning a null result.

\section{Upper limits} \label{sect_UL}

\subsection{Method}
We now describe the procedure employed to compute the upper limits on
the rate of incoming GW short bursts for a set of possible signal
amplitudes: this procedure is separately applied to  each of the 5
subperiods in which the entire observation time $T$ was segmented.
These results are then combined and an overall 95 \% bayesian upper
limit is determined at each signal amplitude.

We remark that, when we compute the upper limit (UL) for a given GW
amplitude, we are assuming  the hypothesis that only signals of that
very amplitude could reach the Earth.
This means that each point in a UL curve is independent of any other
point, and its determination can be independently optimized.  

The handles we have for this optimization are, just as in the
coincidence search previously described, the thresholds to be applied
to the data: varying the thresholds allows us, in turn, to change:
\begin{itemize}
\item the background, i.e. the rate $r_0$, or the mean total number
$\mu_0= r_0 \cdot T$, of accidental coincidences. These are, as before,
estimated with the time-shifted data (sect.\ref{sect_acc}).
\item the efficiencies $\varepsilon$, as computed with the software
injections (sect.\ref{sect_eff}).
\end{itemize}
The output of this search is the estimated maximum rate $r$ of
{\em incoming} GW signals, at any signal amplitude $h_{rss}$ or,
equivalently, the total number  of {\em detected} GW signals
$\mu= \varepsilon \cdot  r \cdot T$.
The optimization consists in choosing the thresholds
potentially capable of producing the best, i.e. lowest, upper limit
$r(h_{rss})$.

We note that the optimization procedure is different from that
employed in sect.\ref{sect_coinci} for the coincidence search: in
that case we looked, at each amplitude, for the thresholds that would
yield the best efficiency for a given (0.1 events) background of
accidentals. Here, not being tied to a pre-fixed value of accidentals,
we can choose the pair  efficiency-background that optimizes our result.

\subsection{The relative belief updating ratio \RO} \label{ULdet}
The quantity we need to compute and optimize, for each value of assumed
signal amplitude, is the {\it relative belief updating ratio \RO},
i.e. the ratio of the likelihood $P(\mu_0+\mu,N)$  that the $N$
coincident events found be due to the presence of a given number
$\mu$ of GW events, to the likelihood $P(\mu_0,N)$ of a mere accidental
background. 

By assuming, as usual, that the number $N$ of coincidences found obeys
the Poisson statistics, we then can write the likelihood
in the presence of a rate $r$ (corresponding to a detectable number
$\mu$) of GW events as:
\begin{equation}
P(\mu_0+\mu,N) = \frac{(\mu_0+\mu)^N e^{-(\mu_0+\mu)}}{N!}
\label{eq_RBUR}
\end{equation}

The same relation, with $\mu=0$, describes the likelihood $P(\mu_0,N)$
of  mere background. Therefore, the relative belief updating ratio
can be written, in terms of our parameters, as:
\begin{equation}
{\mathbf R}(r) =\frac{(\mu_0+\mu)^N e^{-\mu}}{\mu_0^N} 
= \left(1 + \frac{\varepsilon r}{r_0}\right)^N e^{- r \varepsilon}e^{- T}
\label{eq_R}
\end{equation}

The determination of the maximum rate $r(h_{rss})$ of GW, reaching the
Earth with a given amplitude, requires the elaborate procedure
outlined below. For sake of clarity, we describe it for a fixed value
of $h_{rss}$, implying that it is then repeated for all amplitudes of
interest.

Requiring a $95 \%$ confidence limit means finding the particular
value $r^*$ such that $\mathbf{R}(r^*) = 0.05$. Eq. \ref{eq_R} shows
that {\RO} also  depends on other parameters, so that an optimization
is possible: we can vary  our $handles$  (i.e. the thresholds)
till we find the minimum value among all $r^*$.
The functional dependence of  \RO  ~ on the thresholds is due to two
competing effects:
by raising the thresholds, we decrease both $r_0$ (that decreases \RO)
and $\varepsilon$, i.e. $\mu$ (that increases \RO). The result 
can't be analytically predicted, and a numerical search should be done
by varying the thresholds.
Actually, part of this work has already been done in computing the
ROCs (see sect.\ref{ROC}): we found there the optimal efficiency for
each number (or rate) of accidentals. 
This simplifies our search: rather than probing the entire
$(r_0, \varepsilon)$ plane, we only need to compute
$\mathbf {R}(r| r_0, \varepsilon)$ along the ROC curve corresponding
to the $h_{rss}$ considered, as in fig.\ref{fig_roc_coi}.
We now have a family of curves \RO(r) depending on one parameter: the
$(r_0, \varepsilon)$ pair. Of all these curves, we select the one
where the relation $\mathbf {R}(r^*)=0.05$ is achieved with the lowest
value $r^*$ .

One last ingredient is missing for this calculation: the total number
$N$ of coincidences: it is provided by the "on-time" analysis
that cannot be performed before setting all the search parameters.
We have then implemented the following workaround: for each pair
$(\mu_0, \varepsilon)$ from the ROC. We solve for $r^*$ the relation
$\mathbf{R}(r^*)=0.05$ at all possible values of $N_{acc}$. We then
compute the weighted average of these $r$'s:
\begin{equation}
\bar{r^*} = \sum_{N_{acc}%=0} ^{\infty
} r^*(N_{acc}) \cdot P(\mu_0,N_{acc})
\end{equation}
In practice, the sum is truncated when 
\begin{equation}
\sum_{N_{acc}}^M P(\mu_0,N_{acc}) \geq 1 - 1\E{-10}
\end{equation}

Finally, we choose the threshold pair that yields the minimum
$\bar r^*$, i.e. that minimizes the expected UL based on our
efficiency and accidental rate. The search is then repeated at a
different value of GW amplitude, until the curve $r(h_{rss})$ is
traced.

The entire calculation is an "a priori" procedure, performed without
any knowledge of the {\it on time} coincidences.

\begin{figure*}
\includegraphics[width=7cm,height=5cm]{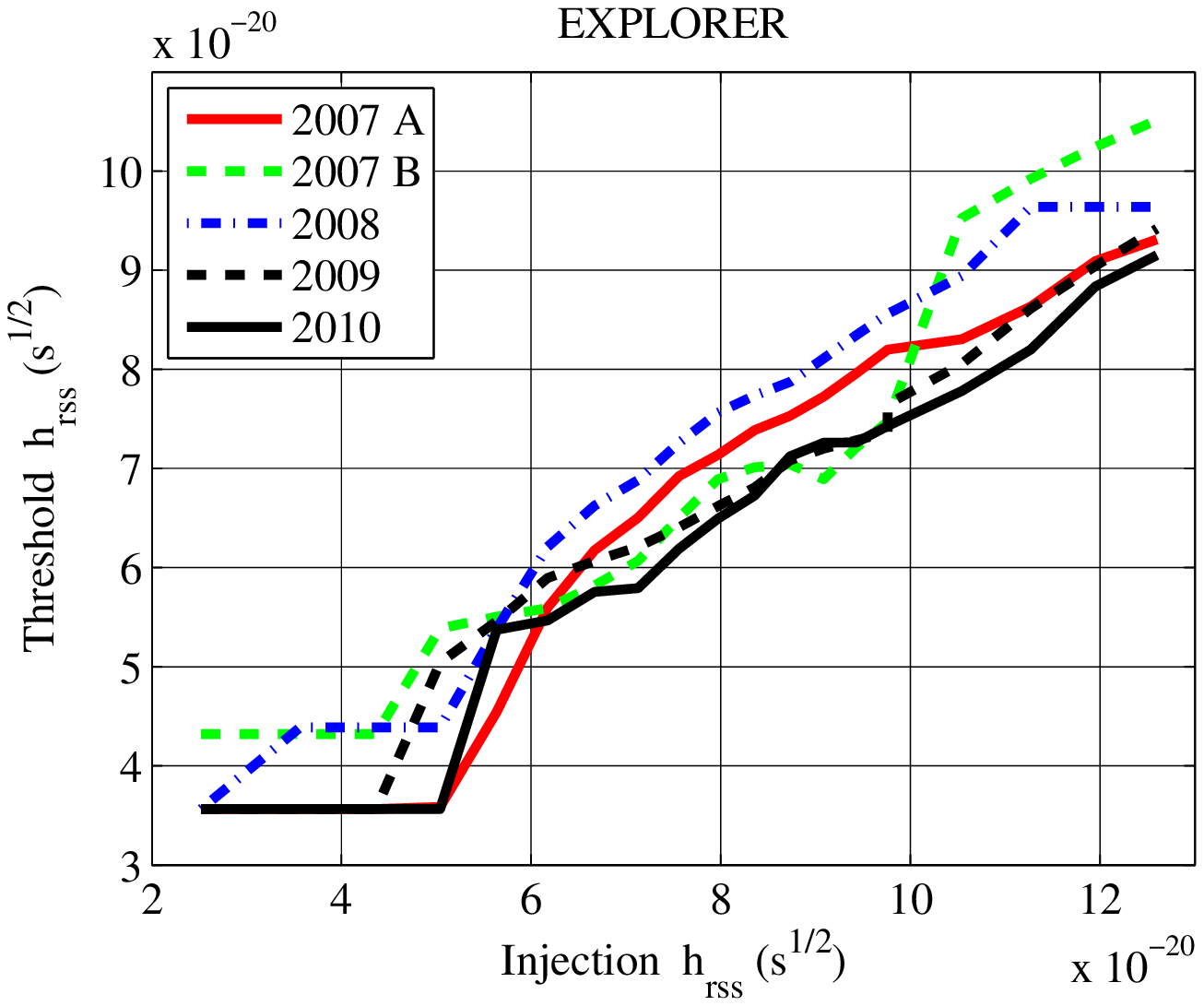}
\includegraphics[width=7cm,height=5cm]{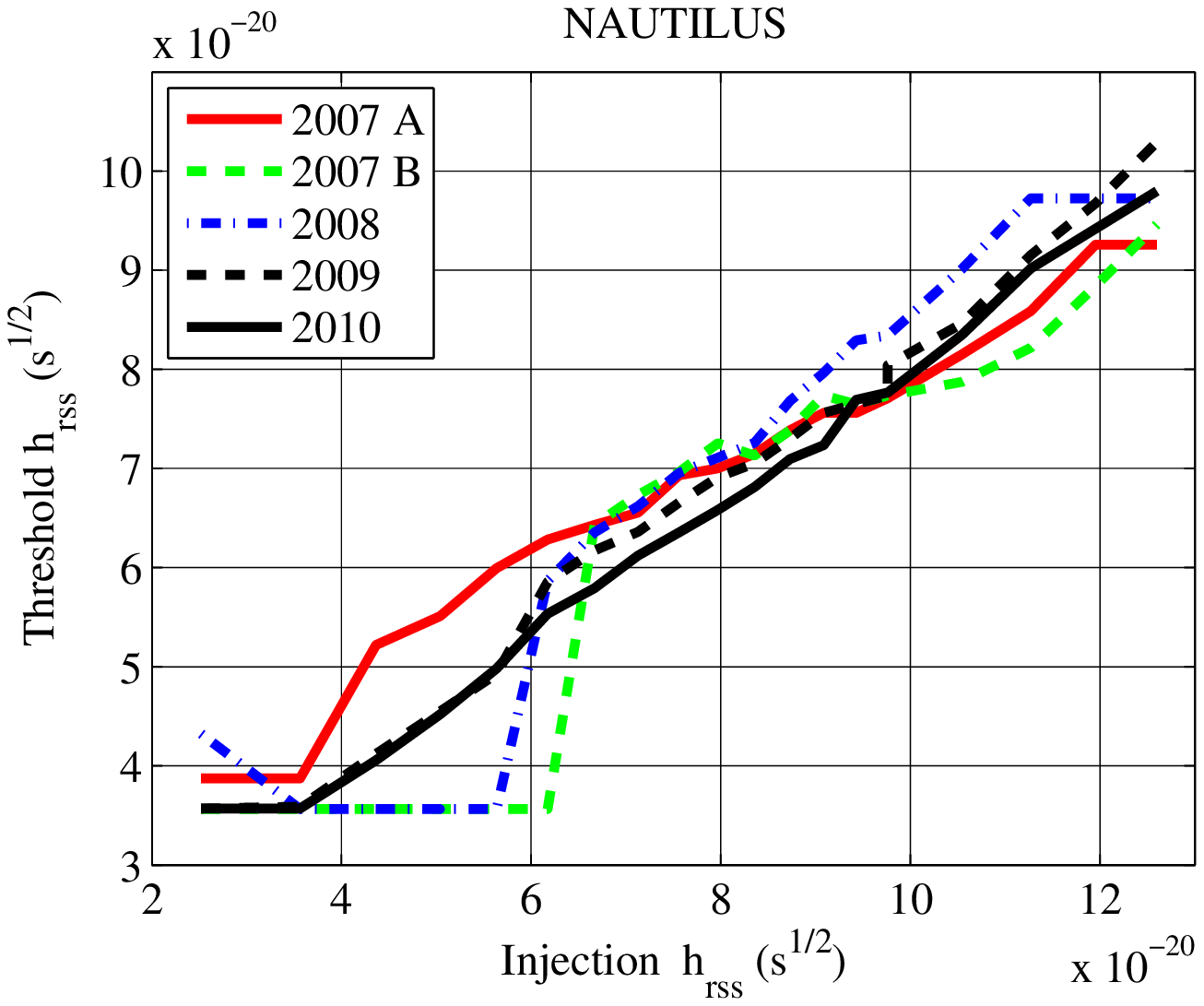}
\label{fig_rc555}
\caption{Thresholds of Explorer and Nautilus resulting from
the UL optimization.}
\vspace*{5mm}
\includegraphics[width=7cm,height=5cm]{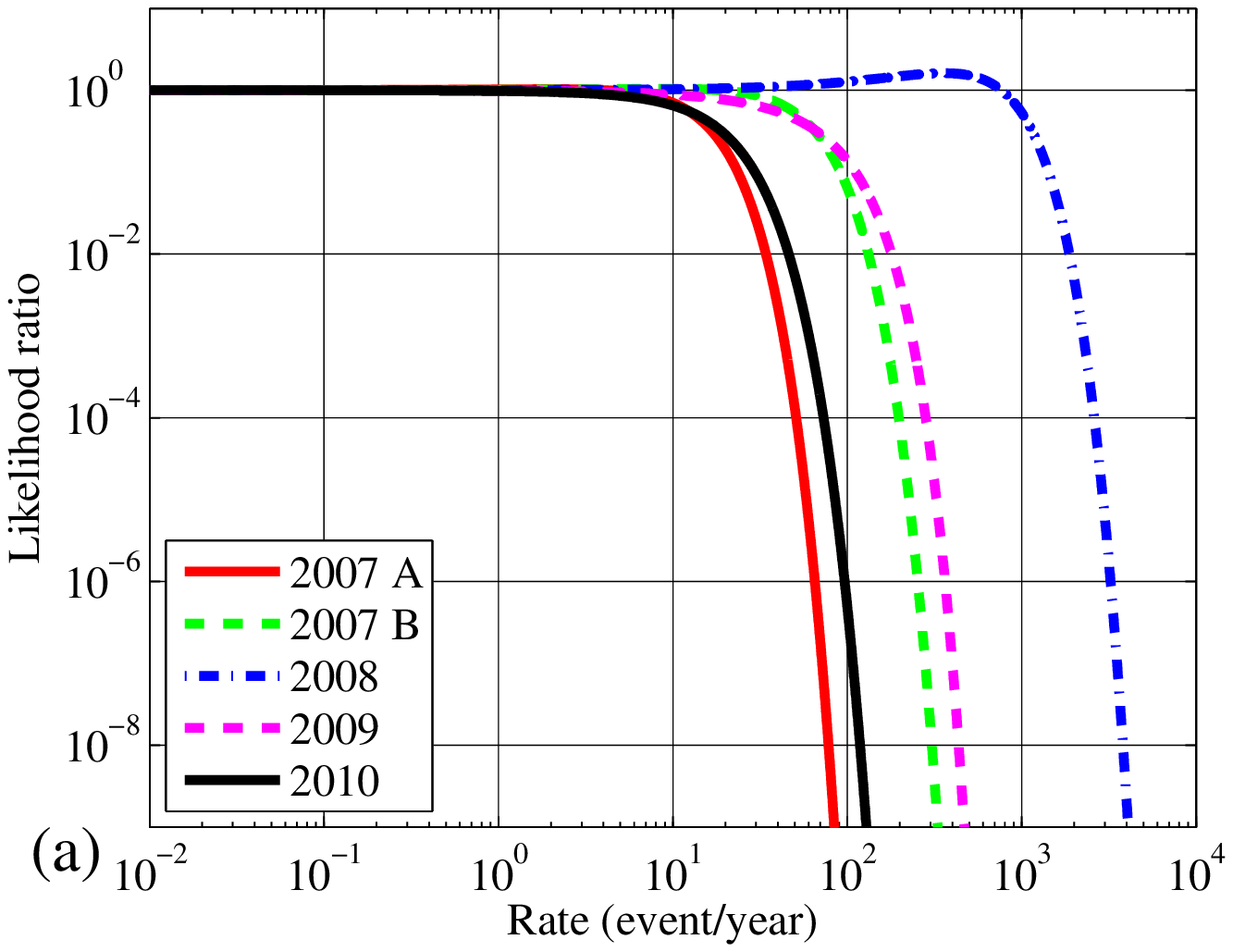}
\includegraphics[width=7cm,height=5cm]{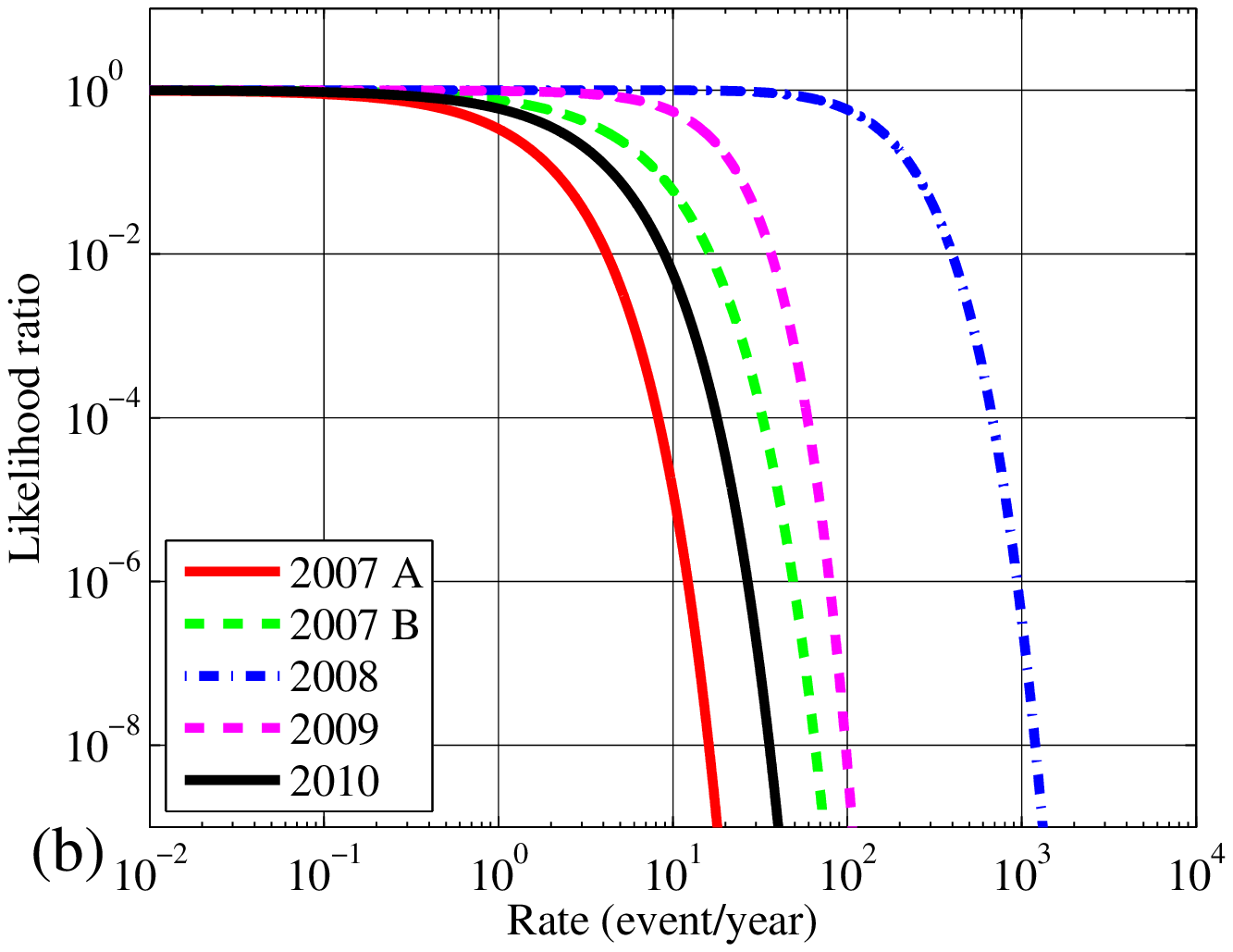}
\label{fig_rc51}
\caption{\RO-curves at $h_{rss }= 5.63\E{-20} $ (a) and  $7.97\E{-20}$
(b) $s^{1/2}$ in the five sub-periods of data taking}
\vspace*{5mm}
\includegraphics[width=7cm,height=5cm]{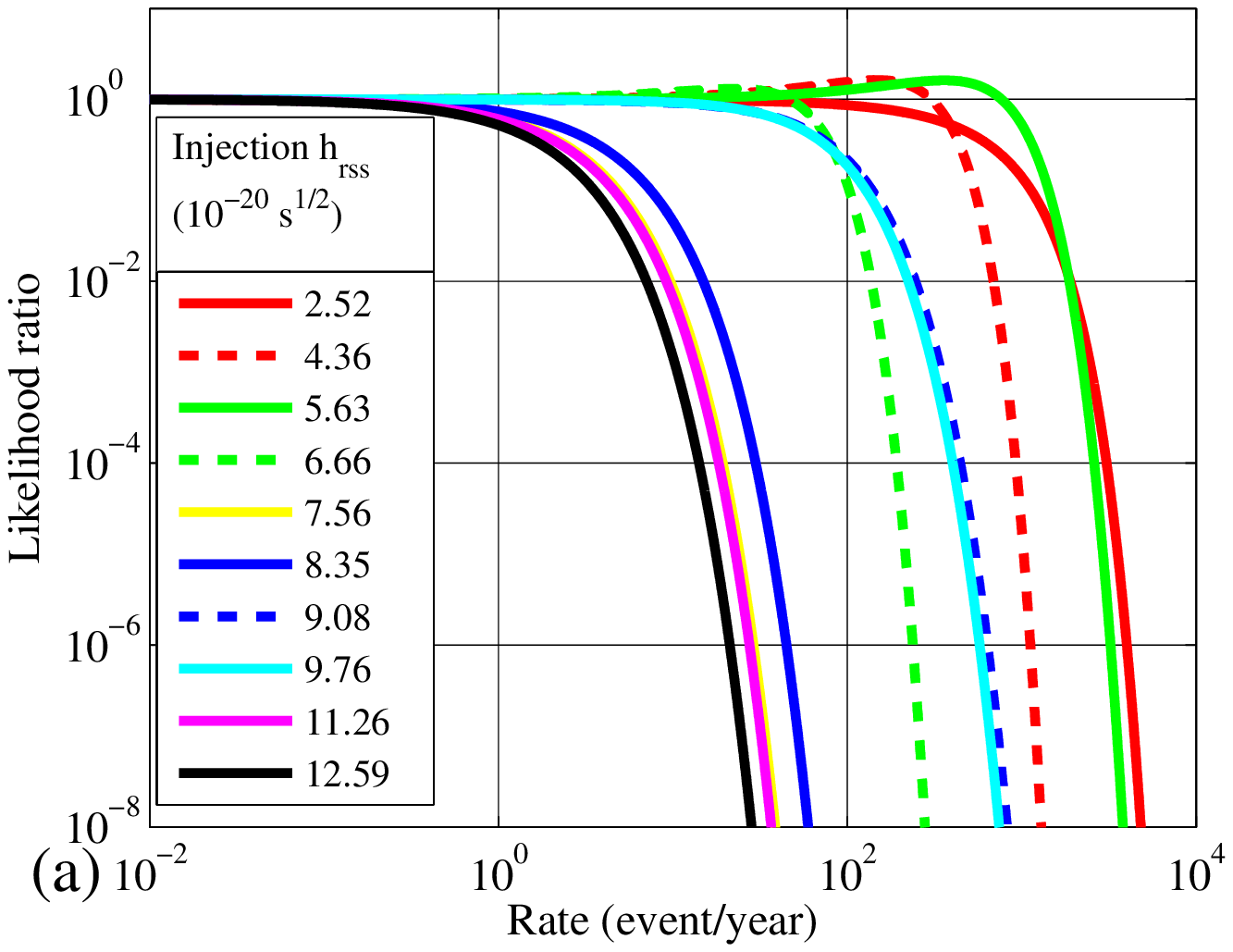}
\includegraphics[width=7cm,height=5cm]{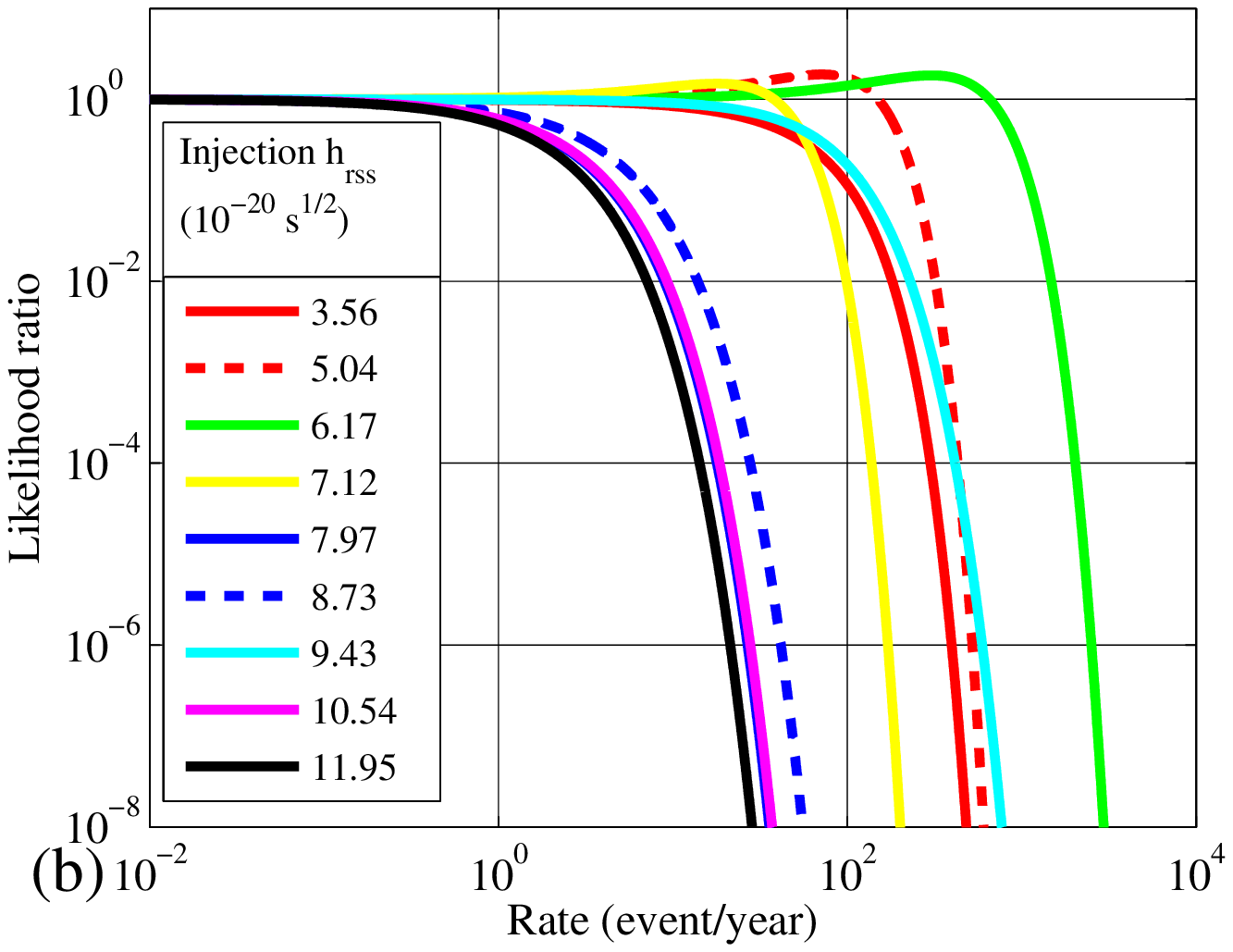}
\label{fig_rctot}
\caption{Total \RO-curves at different values of injected $h_{rss}$.}
\end{figure*}

\subsection{Upper Limit evaluation}
The above procedure was applied to find the optimal threshold pairs
in the 5 time subperiods and for 19 different injection amplitudes
\footnote{Beside the 10 values of injected amplitude used in the
coincidence search, we also explored the small signal region repeating
the injection at 9 additional amplitudes in the range
$[2.52 \div 7.56]\E{-20} s^{1/2}$.} Only at this point, we
had the right to "open the box" and find the {\it on time}
coincidences.

Figure 8 shows the thresholds
for Explorer and Nautilus resulting from the UL optimization.

Figures 9a and 9b show, as an example, the
\RO(r) curves at $h_{rss}^{inj} = 5.63\E{-20}$ and
$7.97\E{-20} s^{1/2}$ in the five subperiods, as well as the total
\RO(r), obtained by multiplying the curves of all subperiods.
In fig.10a and 10b we report the total
\RO(r) computed at various $h_{rss}^{inj}$.

The standard procedure, at this point, calls for evaluation of the UL
as the product of \RO(r) and a {\it prior}, containing all our
previous knowledge on ULs. The best prior is, in principle, a
combination of all previously computed ULs (e.g. \cite{igec3}, 
\cite{ifo}). However, this cannot be used, due to
different meanings and methods of these ULs, as discussed in the next
section. We are then left with the choice of a purely theoretical
{\it prior}: as we make no assumptions on the source location,
polarization, sky distribution etc., it is reasonable to assume a flat
{\it prior}. Our  95\% UL is then simply obtained by picking the value
of $r$ that yields \RO=0.05.

\begin{figure}
\includegraphics[width=8cm,height=7cm]{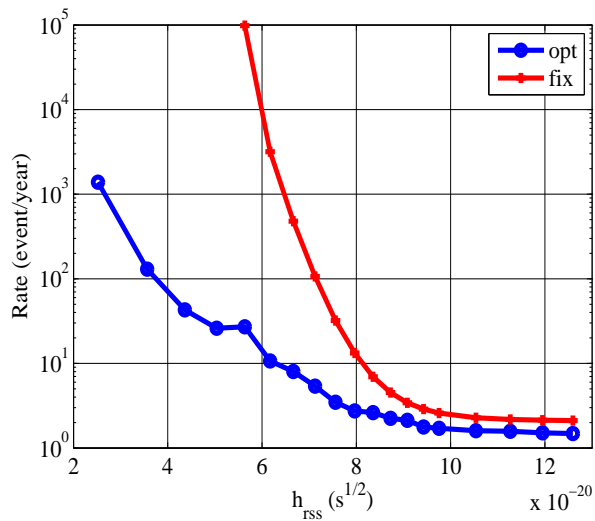}
\caption{Comparison between the upper limit computed with the optimized
procedure described in the text (opt) and with a procedure with
fixed thresholds, those defined by the coincidence search (fix).
There is an evident large gain at low amplitude.}
\label{fig_noopt}
\end{figure}

Figure \ref{fig_noopt} shows the {\it on time} results for the UL
at different values of $h_{rss}$.
The statistical uncertainties on the determination of efficiencies
and accidentals ($\mu_0$) causes on the evaluation of the ULs a
relative error at 1 $\sigma$ ranging from 4\% to 0.4\% for $h_{rss}$
going from the lowest to the highest value, respectively.
The various parameters entering the final evaluation of the UL,
namely the number of {\em on-time} coincidences $N_{coi}$, the
estimated accidentals $\mu_0$ and the computed efficiences, are
reported in tab.V for all the values of injected
$h_{rss}$ and all five subperiods.

\begin{table*}
\centering
\begin{tabular}{||c||c|c|c||c|c|c||c|c|c||c|c|c||c|c|c||}
\hline\hline
 $*1\E{-20}$ & \multicolumn{3}{|c||}{2007 A} & \multicolumn{3}{|c||}{2007 B}
 & \multicolumn{3}{|c||}{2008} & \multicolumn{3}{|c||}{2009}
 & \multicolumn{3}{|c||}{2010}\\
 & & & & & & & & & & & & & & & \\[-4mm]
 $h_{rss}$ & $N_{coi}$ & $\mu_0$ & $\varepsilon$ & $N_{coi}$ & $\mu_0$ & $\varepsilon$
 & $N_{coi}$ & $\mu_0$ & $\varepsilon$ & $N_{coi}$ & $\mu_0$ & $\varepsilon$
 & $N_{coi}$ & $\mu_0$ & $\varepsilon$ \\
\hline\hline
 & & & & & & & & & & & & & & & \\[-4mm]
 2.52 &  565 &  553.9 &  .0245 &   66 &  63.04 &  .0179 &  639 &  663.2 &  .0159 &  283 &  303.4 &  .0193 &   94 &  103.2 &  .0161 \\
 3.56 &  565 &  553.9 &  .1396 &   66 &  63.04 &  .0868 &  638 &  663.2 &  .0910 &  273 &  295.5 &  .1909 &   94 &  103.2 &  .2227 \\
 4.36 &   95 &  85.67 &  .1432 &   66 &  63.04 &  .2139 &  638 &  663.2 &  .2272 &  109 &  130.8 &  .3536 &   40 &  45.68 &  .4378 \\
 5.04 &   57 &  48.94 &  .2532 &   32 &  23.92 &  .2612 &  638 &  663.2 &  .3784 &   30 &  37.01 &  .3699 &   18 &  24.30 &  .6078 \\
 5.63 &   20 &  18.74 &  .3090 &   26 &  21.31 &  .3864 &  390 &  408.2 &  .4192 &   15 &  15.95 &  .4611 &    5 &  3.856 &  .5261 \\
 6.17 &    5 &  7.177 &  .3600 &   25 &  19.89 &  .5002 &  111 &  97.72 &  .3133 &    1 &  3.208 &  .4180 &    2 &  1.536 &  .6222 \\
 6.67 &    2 &  3.588 &  .4339 &    7 &  6.881 &  .4406 &   46 &  41.23 &  .3354 &    1 &  1.457 &  .5283 &    1 &  .7959 &  .7130 \\
 7.13 &    1 &  2.162 &  .5288 &    5 &  4.439 &  .5068 &   30 &  25.37 &  .4041 &    0 &  .9320 &  .6522 &    1 &  .5365 &  .8032 \\
 7.56 &    0 &  .7643 &  .5472 &    3 &  2.930 &  .5518 &   19 &  12.93 &  .4341 &    0 &  .5226 &  .7189 &    0 &  .3088 &  .8393 \\
 7.97 &    0 &  .5404 &  .6346 &    2 &  1.890 &  .5836 &    8 &  8.396 &  .4946 &    0 &  .3342 &  .7762 &    0 &  .2028 &  .8728 \\
 8.36 &    0 &  .3344 &  .6941 &    1 &  1.935 &  .6806 &    7 &  6.411 &  .5696 &    0 &  .2437 &  .8347 &    0 &  .1403 &  .9019 \\
 8.73 &    0 &  .2242 &  .7465 &    1 &  1.646 &  .7292 &    3 &  4.395 &  .6091 &    0 &  .1643 &  .8606 &    0 &  .0749 &  .9031 \\
 9.09 &    0 &  .1503 &  .7892 &    1 &  1.401 &  .7623 &    2 &  3.095 &  .6403 &    0 &  .1297 &  .8899 &    0 &  .0589 &  .9289 \\
 9.43 &    0 &  .1063 &  .8281 &    1 &  1.369 &  .8166 &    0 &  2.122 &  .6620 &    0 &  .1079 &  .9126 &    0 &  .0477 &  .9497 \\
 9.76 &    0 &  .0712 &  .8526 &    1 &  1.248 &  .8491 &    0 &  1.805 &  .7164 &    0 &  .0787 &  .9217 &    0 &  .0387 &  .9650 \\
10.54 &    0 &  .0491 &  .9257 &    0 &  .2660 &  .7679 &    0 &  1.024 &  .7810 &    0 &  .0545 &  .9548 &    0 &  .0223 &  .9834 \\
11.27 &    0 &  .0296 &  .9537 &    0 &  .1186 &  .8439 &    0 &  .5361 &  .7897 &    0 &  .0387 &  .9652 &    0 &  .0135 &  .9892 \\
11.95 &    0 &  .0140 &  .9647 &    0 &  .0594 &  .8952 &    0 &  .5361 &  .8967 &    0 &  .0305 &  .9755 &    0 &  .0084 &  .9919 \\
12.60 &    0 &  .0113 &  .9790 &    0 &  .0265 &  .9210 &    0 &  .5361 &  .9455 &    0 &  .0254 &  .9821 &    0 &  .0068 &  .9959 \\
 & & & & & & & & & & & & & & & \\
\hline\hline
\end{tabular}
\label{tab_carper}
\caption{Characteristic parameters computed in order to evaluate the
\RO$(r)$ curves in the 5 subperiods:  $N_{coi}$ is the $on~ time$
number of coincidences detected,  $\mu_0$ is the estimated average
background and   $\varepsilon$ is the efficiency of detection at each
particular value of $h_{rss}$}
\end{table*}

\begin{figure}
\includegraphics[width=8cm,height=7cm]{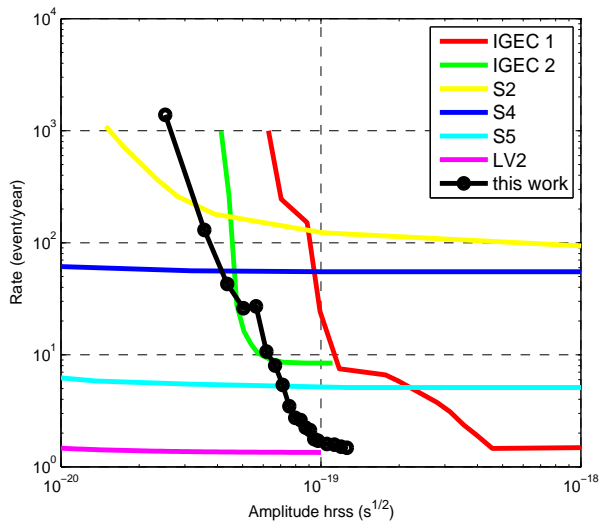}
\caption{Comparison between our $95\%$ Upper Limit and previously
published results. All LIGO and Virgo results, and in particular the solid curve (LV2),
refer to $90\%$ ULs.}
\label{fig_confro}
\end{figure}

\section{Discussion and conclusions}
\subsection{Comparison with other experiments}

Fig.\ref{fig_confro} compares our present results with some other
published in the past: the curve labeled "IGEC 1" is from \cite{igec2},
"IGEC 2" is from \cite{igec3}, "S2" is the UL for 1 ms gaussian
pulses from the LIGO S2 run\cite{LIGOS2}, "S4","S5" and "LV2" are for
Q=9 sin-gaussian pulses at 1053 Hz from \cite{LIGOS4},\cite{LIGOS5}
and \cite{ifo}.

In the present calculation of UL there are several choices that make it
difficult to compare with previous results.

If we consider the above cited upper limits released by the Ligo
Scientific Collaboration (LSC), we can see
that no specific optimization was carried out for the UL: all the
analysis parameters, and in particular the thresholds, were set for
the coincidence search. The number of {\em on-time coincidences} found
was directly used to compute the UL for that run. We show in
fig.\ref{fig_noopt} what the result of such an analysis strategy would
be on our data: the curve labeled "opt" represents the optimized UL
computed in the previous section while the curve labeled "fix" is the
result we would obtain with the thresholds determined in the
coincidence search. The improvement in sensitivity, especially at low
amplitudes can be clearly seen.

A more relevant issue regards the meaning of the variable $h_{rss}$,
i.e. the abscissa of the UL plot. In the two IGEC searches
\cite{igec2,igec3} the efficiency of the detectors was not
considered, and the UL was plotted vs the amplitude threshold used in
the coincidence search. What the curve really meant was then the UL on
GW rates $detectable$ by the observatory with that threshold, rather
than the $incoming$ rate. For a fair comparison, at least the
efficiency of the detectors should be considered: for a given
amplitude $h_0$, when the threshold is set at that very amplitude
$h_0$, the efficiency is roughly $(1/2)^n$  where $n$ is the number
of detectors. The IGEC UL values should therefore be increased by at
least a factor 4 to convert to an $incoming$ rate.

On the other hand, the LSC includes in its analysis a model describing
the distributions of the GW incoming signals, by folding into the
calculation a factor $f(\theta, \phi) \le 1$ to account for isotropic
direction and random polarization of the wave. The $h_{rss}$ in LIGO's
UL is then the maximum amplitude detectable by the observatory.
Besides, they computed a $90 \%$ UL, while ours is at $95 \%$:
according to eq.4.1 of \cite{ifo}, in order to convert their
ULs at 95\%, they should be increased by a factor $\simeq 1.3$.
It's clear that the combination of these two differences, one of
which would lead to a decrease of the UL, the other one to an increase,
would anyhow not change much the comparison with our results.

\subsection{Does the sub-periods segmentation pay off ?}
It is reasonable to question whether a unique search over the entire
observation period $T$ would yield a similar or better result with
respect to our choice of segmenting the analysis in 5 sub-periods.
In other terms, how does the global $\mathbf{R}_G(r)$ compare vs the
product  $\mathbf{R}_T(r)=\prod \mathbf{R}_i(r)$ of 5 separate update
ratios? 
An exact answer can be given in the simple case where we segment T in
two subperiods having the same characteristics, namely accidentals
$\mu_0$ and efficiency $\varepsilon$, assumed constant.
In this case, the global (one period) \RO ~is given by
eq.\ref{eq_RBUR}, while the product of the two \RO's for the
subperiods $T_1, T_2$ is:
\begin{equation}
\mathbf{R}_T(r) = \prod_{i=1}^{2}
\frac{(\mu_i+\mu_{0,i})^{N_i} e^{-\mu_i}}{\mu_{0,i}^{N_i}}
\label{eq_RT}
\end{equation}

As $\mu_0=\mu_{0,1}+\mu_{0,2}$  (and same for $\mu$)
and $N=N_1+N_2$,  one can expand eq.\ref{eq_RT}, and prove 
\RO$_T\equiv$~\RO$_G$,  for any choice of  $T_1, T_2$, $N_1,N_2$.

For the more general case of two non homogeneous subperiods, although
we lack an algebraic proof, extensive numerical investigation has shown
that we should always expect \RO$_T<$\RO$_G$.

\subsection{Conclusions}

In this paper we analyzed 3 years of almost continuous data from the
two resonant gravitational wave detectors Explorer and Nautilus.
The period examined spans from the end of the IGEC2 four-detector
analysis to the decommissioning of Explorer.
Both the search for coincidences with low false alarm rate and the
evaluation of the upper limit have been performed employing a novel
type of analysis, with optimization of the thresholds of each detector
separately for each intermediate task. This method has proven
successful in obtaining better results (see for instance
fig.\ref{fig_noopt}) as well as for handling non stationarities in the
detectors behavior. As an example, we recall the noisy period of 2007B:
in a search with the usual procedure, that period would be discarded,
or its large number of events would negatively affect the statistics
of the remaining, better data. In our case, the optimized procedure
automatically takes care of the higher noise and reduces the weight of
that period on the final results. Indeed, its contribution to both the
coincidence search and the upper limit evaluation is hardly noticeable.

The upper limit computed on the basis of our data cannot compete with
those of the more sensitive interferometric detectors, that extend down
to much smaller $h_{rss}$ values.
Nevertheless, the length of our data collection lead us to expect that
we could improve upon the UL set by LIGO S5, at amplitudes
of the order of $h_{rss} \sim 10^{-19} s^{1/2}$: indeed we did obtain a
better UL than other previously published.
However, while in the process of analyzing our
data, a new, improved UL was released by the LSC-VIRGO collaboration: 
combining the data of the S5/VSR1 and S6/VSR2-3 runs, the extended data
taking allowed them to set a better limit also at higher amplitudes.

The procedure detailed here could be profitably used in future
searches, where better
sensitivity of the detectors would yield even more significant ULs.
Infact, we demonstrated (see fig.\ref{fig_noopt}) that this procedure
grants a substantial improvement in the evaluation of the Upper
Limit, up to two orders of magnitude at low amplitudes, with respect to
the standard way of computing it.

\section*{Acknowledgments}
Explorer ceased taking data on June 11th, 2010, concluding an activity
that spanned over a quarter of a century. We take the
opportunity of this last scientific paper produced with its data to
thank CERN as an institution for hospitality and provision of
facilities, first and foremost the cryogenic fluids. We also thank the
numerous CERN officials and staff that provided, over such a long
period, with logistics and with daily as well as emergency help.\\
We are grateful to the technicians (too numerous to name them all) that,
over the years, helped us maintaining both detectors in operations and,
to these days, still work to keep Nautilus on the air.

\bibliographystyle{amsplain}

\end{document}